\documentclass[twoside,twocolumn,9pt]{article}
\usepackage{extsizes}
\usepackage[super,sort&compress,comma]{natbib} 
\usepackage[version=3]{mhchem}
\usepackage[left=1.5cm, right=1.5cm, top=1.785cm, bottom=2.0cm]{geometry}
\usepackage{balance}
\usepackage{mathptmx}
\usepackage{sectsty}
\usepackage{graphicx} 
\usepackage{lastpage}
\usepackage[format=plain,justification=justified,singlelinecheck=false,font={stretch=1.125,small,sf},labelfont=bf,labelsep=space]{caption}
\usepackage{float}
\usepackage{fancyhdr}
\usepackage{fnpos}
\usepackage[english]{babel}
\addto{\captionsenglish}{%
  
}
\usepackage{array}
\usepackage{droidsans}
\usepackage{charter}
\usepackage[T1]{fontenc}
\usepackage[usenames,dvipsnames]{xcolor}
\usepackage{setspace}
\usepackage[compact]{titlesec}

\usepackage{booktabs}

\usepackage{amssymb}
\usepackage{xr} 
\usepackage{url} 

\usepackage{epstopdf}

\definecolor{cream}{RGB}{222,217,201}

\usepackage{todonotes}

\begin{document}

\pagestyle{fancy}
\thispagestyle{plain}
\fancypagestyle{plain}{
\renewcommand{\headrulewidth}{0pt}
}

\makeFNbottom
\makeatletter
\renewcommand\LARGE{\@setfontsize\LARGE{15pt}{17}}
\renewcommand\Large{\@setfontsize\Large{12pt}{14}}
\renewcommand\large{\@setfontsize\large{10pt}{12}}
\renewcommand\footnotesize{\@setfontsize\footnotesize{7pt}{10}}
\makeatother

\renewcommand{\thefootnote}{\fnsymbol{footnote}}
\renewcommand\footnoterule{\vspace*{1pt}%
\color{cream}\hrule width 3.5in height 0.4pt \color{black}\vspace*{5pt}} 
\setcounter{secnumdepth}{5}

\makeatletter 
\renewcommand\@biblabel[1]{#1}            
\renewcommand\@makefntext[1]%
{\noindent\makebox[0pt][r]{\@thefnmark\,}#1}
\makeatother 
\renewcommand{\figurename}{\small{Fig.}~}
\sectionfont{\sffamily\Large}
\subsectionfont{\normalsize}
\subsubsectionfont{\bf}
\setstretch{1.125} 
\setlength{\skip\footins}{0.8cm}
\setlength{\footnotesep}{0.25cm}
\setlength{\jot}{10pt}
\titlespacing*{\section}{0pt}{4pt}{4pt}
\titlespacing*{\subsection}{0pt}{15pt}{1pt}

\fancyfoot{}
\fancyfoot[LO,RE]{\vspace{-7.1pt}\includegraphics[height=9pt]{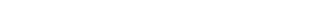}}
\fancyfoot[CO]{\vspace{-7.1pt}\hspace{13.2cm}\includegraphics{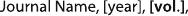}}
\fancyfoot[CE]{\vspace{-7.2pt}\hspace{-14.2cm}\includegraphics{head_foot/RF}}
\fancyfoot[RO]{\footnotesize{\sffamily{1--\pageref{LastPage} ~\textbar  \hspace{2pt}\thepage}}}
\fancyfoot[LE]{\footnotesize{\sffamily{\thepage~\textbar\hspace{3.45cm} 1--\pageref{LastPage}}}}
\fancyhead{}
\renewcommand{\headrulewidth}{0pt} 
\renewcommand{\footrulewidth}{0pt}
\setlength{\arrayrulewidth}{1pt}
\setlength{\columnsep}{6.5mm}
\setlength\bibsep{1pt}

\makeatletter 
\newlength{\figrulesep} 
\setlength{\figrulesep}{0.5\textfloatsep} 

\newcommand{\topfigrule}{\vspace*{-1pt}%
\noindent{\color{cream}\rule[-\figrulesep]{\columnwidth}{1.5pt}} }

\newcommand{\botfigrule}{\vspace*{-2pt}%
\noindent{\color{cream}\rule[\figrulesep]{\columnwidth}{1.5pt}} }

\newcommand{\dblfigrule}{\vspace*{-1pt}%
\noindent{\color{cream}\rule[-\figrulesep]{\textwidth}{1.5pt}} }

\makeatother

\twocolumn[
  \begin{@twocolumnfalse}
{\includegraphics[height=30pt]{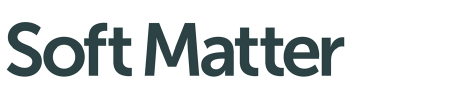}\hfill\raisebox{0pt}[0pt][0pt]{\includegraphics[height=55pt]{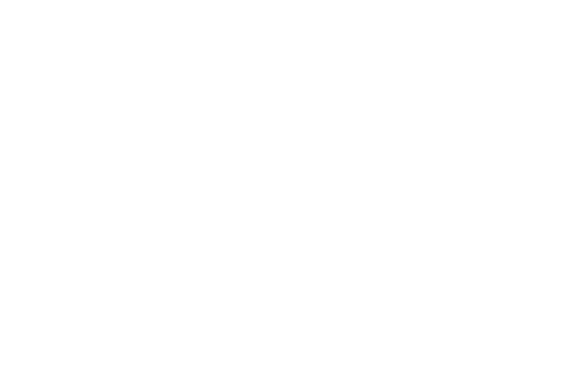}}\\[1ex]
\includegraphics[width=18.5cm]{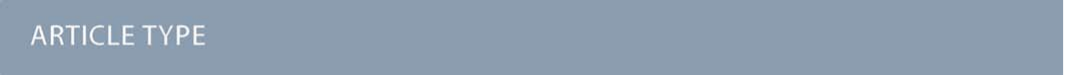}}\par
\vspace{1em}
\sffamily
\begin{tabular}{m{4.5cm} p{13.5cm} }

\includegraphics{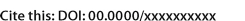} & \noindent\LARGE{\textbf{Screened Colloidal Interactions Alter Network Morphology of Confined Colloidal Drying Patterns}} \\
\vspace{0.3cm} & \vspace{0.3cm} \\

 & \noindent\large{Andreas Andersen Hennig,\textit{$^{a}$} Natalia Kizilova,\textit{$^{b,c}$} Ilaria Beechey-Newman,\textit{$^{a}$} and Erika Eiser$^{\ast}$\textit{$^{a,d}$}} \\

\includegraphics{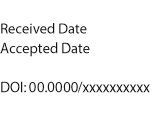} & \noindent\normalsize{Colloidal suspensions that are confined between two closely spaced plates and left to evaporate slowly form maze-like labyrinth patterns in the final stage of evaporation. 
We compare drying patterns formed by colloidal droplets suspended in deionised water and in a 50\,mM NaCl solution, thereby investigating the effect of reducing the electrostatic screening length, and hence modifying the effective interparticle interactions. 
The fingers in these patterns form bifurcations which we may quantify with a skeletonization method.
We find that the junctions generally form thinner fingers towards the center and that the typical bifurcation angles peak around $54^\circ$ when using deionized water and $58^\circ$ when we add small amounts of salt. 
Droplets with screened particles are characterized by bands of transversal fingers before merging into a central non-fingering colloidal aggregate, characterized by the aggregation radius of $r_a=0.74\pm0.14$\,mm. 
Our observations suggest that colloid-colloid interactions can strongly influence the large-scale morphology of the pattern while leaving the local geometry of the individual fingers mostly unchanged. } \\

\end{tabular}

 \end{@twocolumnfalse} \vspace{0.6cm}

] 

\renewcommand*\rmdefault{bch}\normalfont\upshape
\rmfamily
\section*{}
\vspace{-1cm}


\footnotetext{\textit{$^{a}$~PoreLab, Department of Physics, Norwegian University of Science and Technology, H{\o}gskoleringen, Trondheim, N-7491, Norway. E-mail: erika.eiser@ntnu.no}}
\footnotetext{\textit{$^{b}$~Warsaw University of Technology, Institute of Aeronautics and Applied Mechanics, Nowowiejska, 24, Warsaw, 00-665, Poland.}}
\footnotetext{\textit{$^{c}$~V.N. Karazin Kharkiv National University, Department of Applied Mathematics, Svobody sq., 4, Kharkiv, 61-022, Ukraine.}}
\footnotetext{\textit{$^{d}$~Cavendish Laboratory, University of Cambridge, Cambridge, CB3 0HE, United Kingdom.}}


\section{Introduction}
Evaporation of colloidal suspensions can be surprisingly intricate, and offer explanations to everyday phenomena like the famous ``coffee-ring effect", where most of the solid particles in a droplet of coffee heap up in the rim instead of being spread evenly. 
The effect was first explained by Deegan \textit{et al.}\cite{Deegan_etal_1997a}, and holds for sessile droplets on partially wetting surfaces with a pinned contact line. As the droplet evaporates, they show there will be a capillary flow towards the edge, transporting almost all colloidal particles towards the perimeter. In both inkjet printing, and industrial coating and painting, this effect is unwanted, which has lead to efforts to overcome these capillary flows. 
Notably, the coffee-ring effect could be suppressed and even reversed by inducing Marangoni flows that oppose outward capillary flow, by using octane droplets\cite{Hu_Larson_2006a}, completely removing impurities (especially surfactants) from pure water \cite{Hu_Larson_2006a}, using surfactant-like viscous polymer solutions \cite{Seo_etal_2017}, or allowing evaporation into an ethanol-saturated atmosphere \cite{Majumder_etal_2012}, to name just some examples.  

Recently, Beechey-Newman \textit{et al.}\cite{Beechey-Newman_etal_2025} showed that a completely different, maze-like pattern emerges when a strongly confined colloidal suspension is allowed to dry very slowly. 
These patterns form finer and finer fingers of colloid deposits on both the top and bottom surface of the confining cell, due to the crumpling of the crust of colloidal material that develops at the air-water interface of the capillary bridge. The instability that forms these fingers is not yet fully explained, but is most likely due to both thermal forces and hydrodynamic interactions, making the system a challenge to model properly.

Such circularly symmetric fingering patterns are found in a variety of natural phenomena, most notably in viscous fingering and dendritic growth patterns -- both commonly modeled by Diffusion-Limited Aggregation (DLA) \cite{Witten_Sander_1981}. 
Recently, such roughly circularly symmetric patterns were described by their arc-lengths, showing a transition from a stable number of fingers to a power-law relation \cite{Hennig_etal_2026a}. 

Here, we aim to describe the network properties of the maze-like patterns, as the exact structure of colloidal networks may reflect the physical origin of the system \cite{Tsurusawa_Tanaka_2023}.
We show that the skeleton may shed light on the effect the Coulomb repulsion and thus the screening lengths between the colloids in aqueous suspensions.
The paper is structured to first give an overview of drying patterns of colloidal droplets, with the variety of morphologies that are typically observed.
We then describe 24 fingering patterns formed by slow evaporation in confinement as skeletonized networks, where 12 are formed in the same way as described by Beechey-Newman \textit{et al.}\cite{Beechey-Newman_etal_2025, Hennig_etal_2026a} and 12 are formed with a salt concentration of 50\,mM salt in the evaporating droplet. 

\section{Overview of colloidal drying patterns} 
In recent decades the formation of structured sediments in drying colloidal/polymer drops, liquid films \cite{Haidara_etal_2001, Yang_etal_2015, Harris_etal_2007} and at liquid-liquid interfaces has become the subject of numerous experimental and theoretical studies and numerical simulations \cite{Thampi_Basavaraj_2023}. 

A variety of the morphological types from highly ordered monolayers \cite{Park_Moon_2006, Zhang_etal_2016} and uniform sediments \cite{Sommer_etal_2004, Zhong_Duan_2015, Anyfantakis_etal_2015} to concentric/nonconcentric rings \cite{Deegan_etal_1997a, Deegan_2000a, Hu_Larson_2006a, Zhang_etal_2008, Shen_etal_2010, Maheshwari_etal_2008}, discrete clusters \cite{Fang_etal_2002, Jang_Oh_2004}, dendrites \cite{Haidara_etal_2001, Mougin_Haidara_2002, Jang_Oh_2004, Chen_etal_2005a}, viscous-fingering-like \cite{Maheshwari_etal_2008}, wormlike \cite{Liu2020}, fractal \cite{Chen_etal_2005a, Fang_etal_2006}, complex networks \cite{Fang_etal_2002, Maillard_etal_2001}, spinodal \cite{Tang_etal_2002, Seknazi_etal_2019} and labyrinth \cite{Beechey-Newman_etal_2025} structures is determined by concentration, size and shape of the colloidal particles \cite{Zhang_etal_2008}, droplet size \cite{Shen_etal_2010}, solvent type, surface energy and topography of the substrate \cite{Mougin_Haidara_2002}, ambient temperature and humidity, and deposition method. 

Uniform deposition is essential for technological applications like painting, coating, ink-jet printing, etc. The uniformity of sediments can be controlled by changing the physical properties of the particles or by adding a surfactant \cite{Mampallil_Eral_2018, Seo_etal_2017, Kim_etal_2016}. Complex morphologies have been observed in aqueous suspensions of different nanoparticles (NPs) \cite{Haidara_etal_2001, Tang_etal_2002, Jang_Oh_2004} and polymers \cite{Smalyukh_etal_2006, Fang_etal_2006, Maeda_1999, Maeda_2000}.

Among biological liquids and particles, the dried droplets with concentrated virus dispersions \cite{Fang_etal_2002}, DNA solutions \cite{Fang_etal_2002}, DNA-silica colloidal suspensions \cite{Zhang_etal_2008}, native blood and blood plasma \cite{Chen_etal_2016, Jaber_etal_2021, Pal_etal_2020}, collagen solutions \cite{Maeda_1999, Maeda_2000}, and different protein solutions \cite{Yao_etal_2023} exhibited different complex structures. The patterns of accumulated matter and cracks in the dried blood drops are promising for pattern-based medical diagnosis \cite{Chen_etal_2016, Sefiane_etal_2021}. 

Evaporation-driven pattern formation is a self-organized process that allows fast and cheap production of micro/nanostructures in microelectronics and medical engineering devices. Especially evaporative colloidal lithography and microfluidic electronics \cite{Harris_etal_2007, Wang_etal_2018, Dudala_Goel_2024} control the mechanisms for appearance and growth of the evaporative self-assembled structures. 

Self-assembled structures of metal, metal oxide and semiconductor NPs are important for magnetic and optical applications. Catalytic, magnetic, electric and optical properties of such structures are morphology-dependent and differ from the physical properties of their bulk material \cite{Jang_Oh_2004, Lim_etal_2010, Brust_etal_1998, Khatri_etal_2008}. In more complex multicomponent systems, the complexity of the pattern formation at the liquid-solid interface is very important for technical applications due to the competing intermolecular and interfacial interactions between different species and substrates. 

The variety of deposited structures can be arranged in several groups depending on the mechanism of their formation: 

1) Periodic hexagonal cells induced by Benard-Marangoni convection. The evaporation of solvent at the liquid-air interface decreases the surface temperature that produces a temperature gradient orthogonal to the solid surface; the generated Marangoni stresses are essential. The surface tension gradient in the liquid film causes Benard-Marangoni instability that leads to hexagonal patterns upon drying \cite{Nguyen_Stebe_2002, Wang_etal_2001, Maillard_etal_2001}. The stability of the liquid-vapour interface depends on two non-dimensional control parameters, the capillary and Marangoni numbers \cite{Sultan_etal_2005}. 

2) Coffee rings formed by capillary flow \cite{Deegan_etal_1997a}. The pinned contact line (CL) of the evaporating drop produces an outward capillary flow to compensate for the enhanced loss of solvent at the CL. The microflows continuously carry the solute with particles toward the droplet edge that leads to an accumulation of the particles at the perimeter of the drying drop. Typical multiple-ring patterns are produced by the periodical stick-slip motion of the shrinking CL line, which is governed by the competition between pinning forces and surface tension at the CL \cite{Shmuylovich_etal_2002, Adachi_etal_1995, Maeda_1999, Maheshwari_etal_2008}. 

3) Oppositely charged particle-surfactant systems produce homogeneous deposits with a pinned CL; intermediate surfactant concentration almost entirely suppresses the particle charge, making them hydrophobic and promoting particle trapping into the liquid-air interface. This results in a particle skin or crust forming along the droplet interface, which may then be deposited onto the drying surface\cite{Anyfantakis_etal_2015}.

4) DNA may also be transported to the CL, as for the coffee-ring effect, and form a lyotropic liquid crystal (LC) whose structural orientation is concentric or parallel to the interface, instead of showing typical 'molecular combing' or perpendicular orientation observed at a liquid meniscus concentric chain orientations, to minimize the LC elastic energy. During the final stages of drying, the contact line retracts, and the radial stress causes undulations at the rim that propagate inward through the LC and form a periodic zigzag structure\cite{Smalyukh_etal_2006}.

5) Viscous fingering-like patterns at the edge of the drop are induced by fluctuations in the CL pinning and local curvature change and dimple formation at the interface \cite{Takhistov_Chang_2002}. The stagnation of microflows near the CL is responsible for particle accumulation and deposition. When a less viscous fluid is pushing against a more viscous fluid, the miscible viscous fingers appear due to destabilization of localized disturbances \cite{Ben_Demekhin_2002, Maheshwari_etal_2008}. Then the fluid with lower viscosity moves with higher speed and penetrates the more viscous fluid (Saffman-Taylor instability).

6) Spoke-like structures appear due to competition between receding CL and particle delivery by osmotic-driven microflows and deposition. The spokes located perpendicular to the receding CL. In the first stage, the pinned CL causes the outward fluid flow to compensate for solvent loss at the perimeter of the drying droplet. The particles transported by the flow accumulated along the CL. As the CL receded, clusters of particles are pushed inward and continued to grow as more particles are trapped during the receding, and the thicker parts recede more slowly than the thinner parts. These growing clusters provide greater resistance to CL movement and prevented it from depinning at this point. When the CL finally depin in the vicinity of the immobile colloid clusters, the receding CL with higher curvature-dependent evaporation at the spike causes a tangential flow, which leads to further compaction of the clusters and the formation of spikes along the CL. At the last stage, the receding CL can no longer remain pinned on the cluster obstacles, and it moves inward, leaving behind spoke-like structures \cite{Poulard_etal_2003, Yang_etal_2014, Zhang_etal_2008}.

\begin{figure*}[ht!]
\centering
\includegraphics[width=0.9\linewidth]{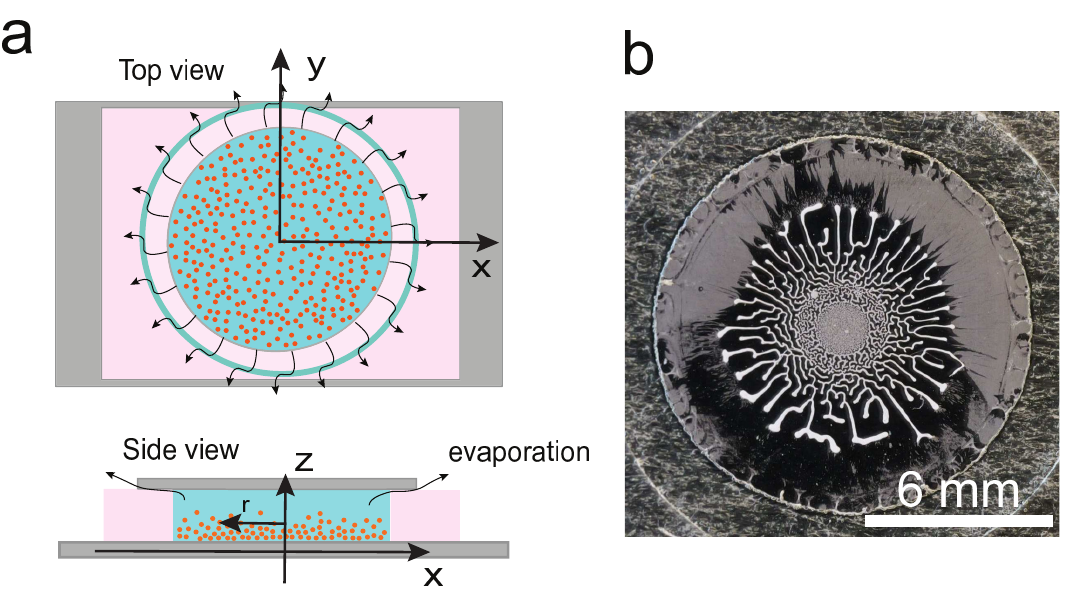}
\caption{(a) Cartoon of the top and side views of the sample cell. (b) Photograph of a fully dried sample (CS8 in Figure \ref{fig:Full_gallery}) showing the three drying stages.}
\label{fig:Sample_preparation}
\end{figure*}

The coffee-ring effect, diffusion-limited aggregation and crystal monolayers remain the most studied systems. But the generation of the complex evaporation-driven morphologies in drying drops (thin films) still harbours open questions \cite{Rabani_etal_2003, Bigerelle_etal_2006}, including those recently studied by Beechey-Newman \textit{et al.}\cite{Beechey-Newman_etal_2025}. In this case evaporation only occurs along the one-dimensional perimeter of the coverslip as shown in Figure\,\ref{fig:Sample_preparation}. The resulting patterns show a complex combination of multi-rings, crystal monolayers, interface instabilities and spoke-like structures. 

\section{Experimental methods}
First, we synthesize TPM (3-(trimethoxysilyl) propyl methacrylate) particles with a diameter of roughly $1.8$\,$\mu$m and zeta potential of about -36\,mV in water using a method based on that described by\cite{VanDerWel_etal_2017}. A detailed description is given by Beechey-Newman \textit{et al.}\cite{Beechey-Newman_etal_2025}. The colloidal solution was washed with deionized water 3 times before use. 
 
Flat, cylindrical cells were prepared by placing a double sided sticky tape with a circular hole (12\,mm diameter) on a clean microscope slide. 30\,$\mu$l of a 1\,wt.\% TPM dispersion was pipetted into the cell, which was then sealed with a cover-slip such that the entire cell was filled. The quasi-sealed sample was then left to slowly evaporate until dry in ambient conditions, which typically takes $8\pm 2$ days. In Figure\,\ref{fig:Sample_preparation} we show a cartoon of the sample cell and a high resolution photograph of a fully dried sample that was taken with a Lumix S5 II camera equipped with a SIGMA 70\,mm 1:2:8 DG Macro $\varnothing49$ macro zoom. 

\begin{figure*}[ht]
\centering
\includegraphics[width=0.90\linewidth]{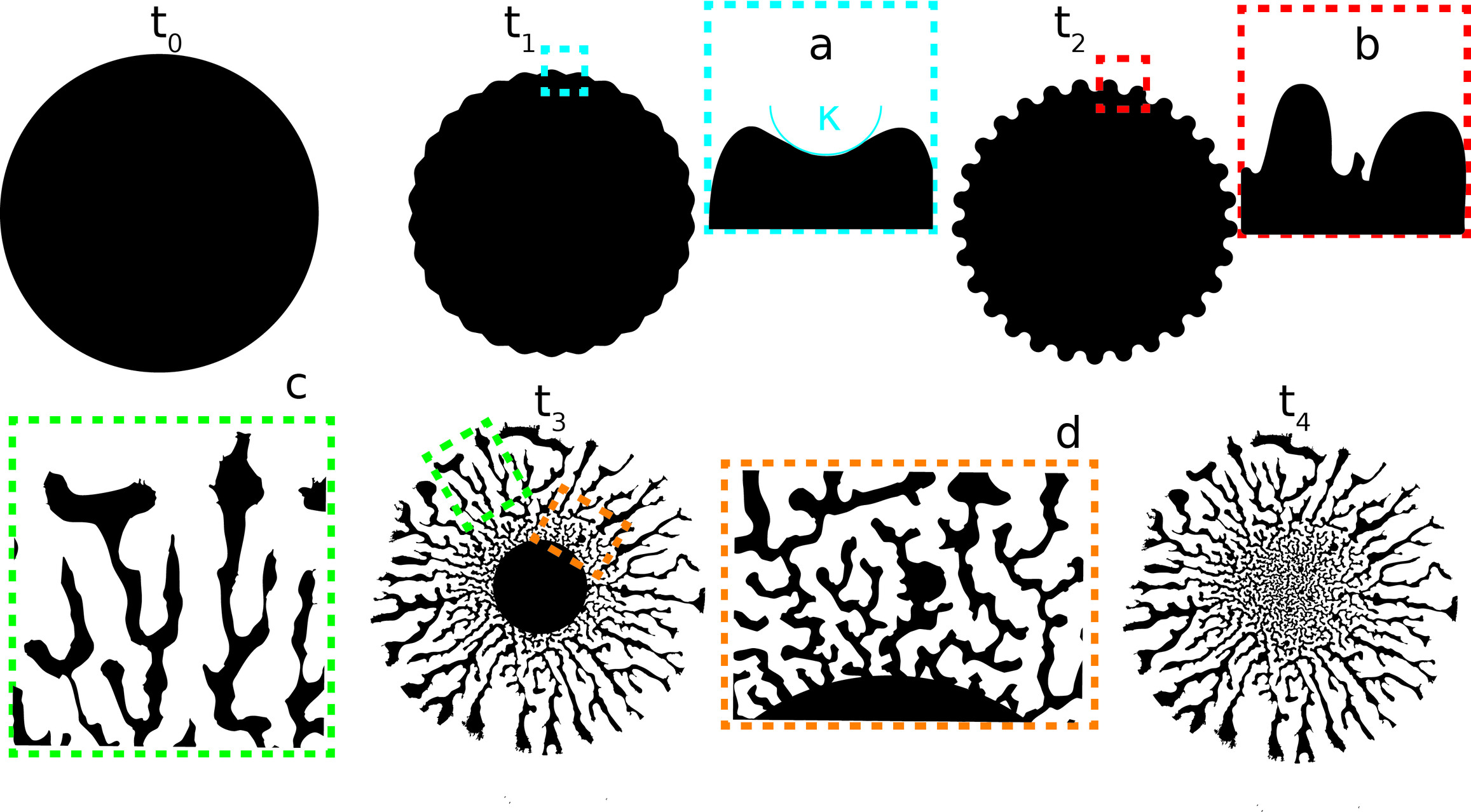}
\caption{Schematic of the last stage of the drying process where the fingering instability occurs, seen from above (top row for illustration only). $t_0$ marks the moment shortly before the fingering instability, when the detached droplet starts to show an accumulation of colloids at the air-water interface. Finger formation starts at $t_1$, while additional fingers start to emerge from the ``inscribed liquid circle'' at $t_2$. The inscribed circle continues to shrink ($t_3$), while further fingers form. $t_4$ show the final self-assembled colloidal pattern (\emph{i.e.} the mass phase) of the image shown in Figure\,\ref{fig:Sample_preparation}b. Once formed the fingers do not reorganize. Inset (a) shows the schematic onset of the formation of dimples with curvature $K$ in the air-water interface. (b): As the fingers form, new fingers emerge in the void spaces in-between. (c): Binarized image of the emerging fingers; the number of fingers remains roughly constant with radius \cite{Hennig_etal_2026a}. (d): At some point, no new fingers form while existing fingers merge to form a labyrinth.}
\label{fig:Physical_Model}
\end{figure*}

We observe three regions that show distinctly different drying behaviors\cite{Beechey-Newman_etal_2025}: Stage\,1 shows the appearance of air bubbles that grow along the perimeter of the cell wall. This stage typically takes up half of the total drying process and ends when all air bubbles have coalesced, leaving the droplet fully detached from the outer wall. Stage\,2 then shows the capillary drop shrinking uniformly and which leaves behind  a monolayer of colloids on both the bottom and top surface. 
In the third and final stage, the monolayer deposition stops abruptly, and the formation of fingers occurs due to an instability in the air-water interface (see Figure\,\ref{fig:Physical_Model})\cite{Beechey-Newman_etal_2025}. 
Here, we only focus on the third stage.

We compare samples under two different solvent conditions: ``standard" samples C1-C12 were prepared in deionized water, while salted samples CS1 to CS12 where prepared with deionized water containing 50\,mM of NaCl.
The presence of ions screens and thus weakens the effective Coulomb repulsion between the colloids dispersed in water, according to classical DLVO theory \cite{israelachvili2011intermolecular,Brochard-Wyart_etal_2019}.

While fingering instabilities occur in all samples, uncertainties in our setup lead to differences in the pattern morphology.
Patterns C1, C6 and C10 are characterized by fairly large, thick fingers on one side of the pattern, which may arise from the fact that the samples were not placed perfectly horizontally, or by an uneven placement of the cover-slip, which may cause a higher evaporation rate at one side.
Pattern C9 have a clear spiral-like structure and C4-C5 become faint towards the center, seemingly running out of colloids to keep the instability to the central point.

Generally, we find that the patterns forming with 50\,mM salt were more symmetric and qualitatively form more similar patterns.
A full gallery of the patterns we analyze is found in Figure \ref{fig:Full_gallery}.

\section{Skeletonization of colloidal fingering patterns}
Complex network patterns found everywhere in nature are the results of determined physical and (bio)chemical processes, which are governed by stochastic processes and instabilities. Therefore, statistical methods play an important role in determining the main physical phenomena underlying the dynamics of the system considered. Classical examples are the bifurcations in blood vessels \cite{Zamir1982, LaBarbera1990}, lungs \cite{Horsfield1985}, tree branches \cite{Eloy2017}, and plant leaves \cite{Kizilova2004, Kolivand2019}.

\begin{figure*}[ht!]
\centering
\includegraphics[width=0.9\linewidth]{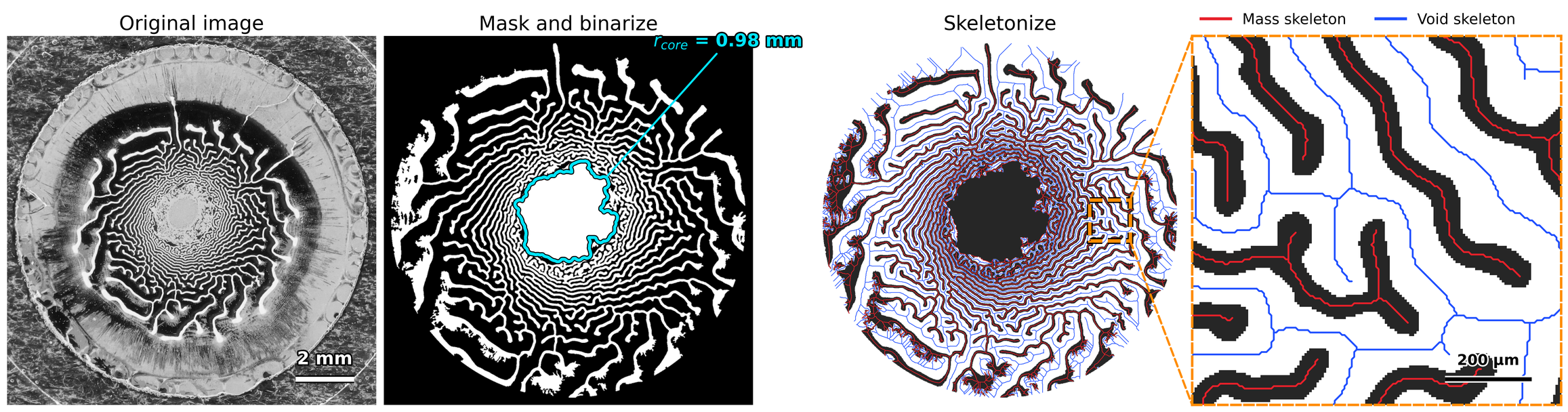}
\caption{Illustration of the image processing on pattern CS10 to obtain the clean skeletons from our original images. The process uses Otsu's algorithm for binarization, and is converted into a 1 pixel thick skeleton. We also skeletonize the voids between the fingers.}
\label{fig:Image_Processing}
\end{figure*}

\begin{figure*}[ht]
\centering
\includegraphics[width=0.9\linewidth]{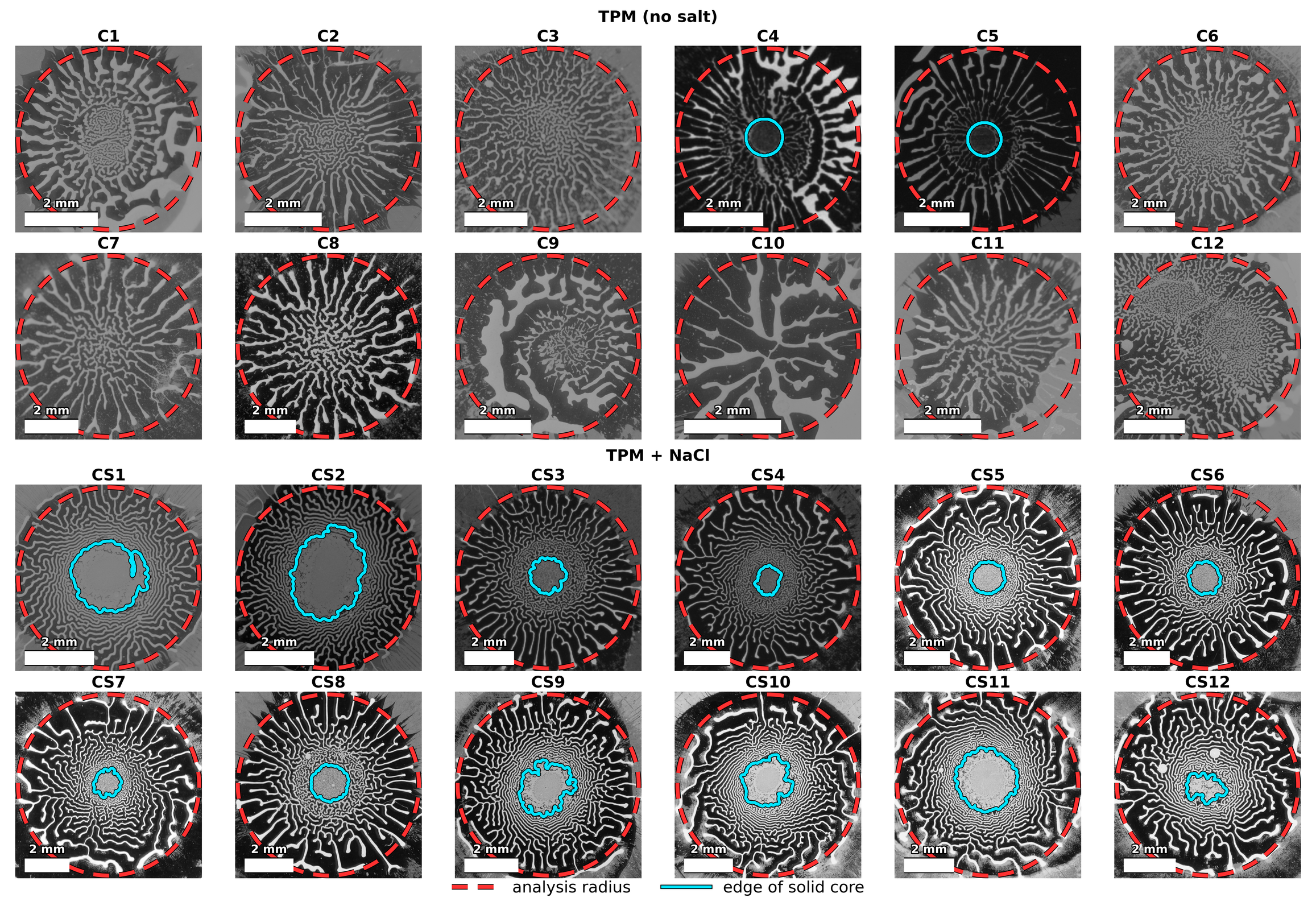}
\caption{Original images we use for this paper, with scale bars of 2\,mm. The pattern's maximal radius is illustrated by a red dashed circle, and the edge of the central aggregate is shown for the salted patterns (CS1-CS12) and 2 of the unsalted ones. To measure the central aggregate, we use the same method as in \cite{beechey2026elasticity}.}
\label{fig:Full_gallery}
\end{figure*}

In Figure \ref{fig:Image_Processing} we show the process of skeletonizing the network structure of our patterns, here starting from the macro-zoom photograph of the sample CS10, containing NaCl. 
We mask the monolayer and void regions (black) while the fingering pattern is set to white. 
This is done as follows: First we isolate the pattern by picking a rough center to create the black circular mask.
Then, we use Otsu's method of binarizing our image \cite{Otsu_1979}, with a visual inspection that filters away the gray monolayer and darker void region from the white finger pattern.
After the image is binarized, we use the Hoshen-Kopelman labeling algorithm \cite{Hoshen_Kopelman_1976} to label all independent clusters, and delete all specks to remove unwanted noise from our images \cite{hennig_2026_18954724}.

We then skeletonize the binarized image using the \texttt{skan} library in Python, which was originally developed by Nunez-Iglesias \textit{et al.}\,to study cytoskeletons in red blood cells \cite{Nunez-Iglesias_etal_2018}.
The algorithm thins the fingers iteratively, until all fingers are 1 pixel in thickness, and is similar to the Fiji/ImageJ Analyze Skeletons plugin \cite{Arganda-Carreras_etal_2010}.
In this work, we will analyze the network in both phases of the pattern: the fingers that are made of the colloids is the mass phase and the spaces between them, which is void of colloids. 

In order to study how the network evolves with the distance to the center of the pattern, we need a robust method of determining the center-point.
One problem of determining the exact center-point of these patterns arise from their disorder. 
The center of mass is the simplest measure of a center, but is clearly skewed by the thick outer fingers occurring in most patterns, as seen in samples C1, C9 and C10 in Figure \ref{fig:Full_gallery}. 
Another method of determining the center-point is to follow the network towards its most connected node, as done in the Strahler method, which was developed to classify the hierarchical degree of branching in river-networks and trees\cite{Strahler1957}. 
However, this method assumes full network connectivity, which is generally not guaranteed.
Therefore, we use the entropy-center method we introduced to determine a center in our roughly radial, disordered, noncontiguous drying patterns\cite{Hennig_etal_2026a}.
This method relies on selecting a window of $N\times N$ relevant center-points, $X_i$, and a range of relevant radii $r\in [r_{\text{min}}, r_{\text{max}}]$.
At each point $X_i$ and radius $r$, we may draw a telegraph signal around the circle, where 1 indicate mass phase and 0 indicate void phase. 
Thus, for each point we may find the average telegraph signal over the selected range of radii, $\mathcal{T}_S(\theta; X_i)$.
We compute the power spectral density of this signal, $G(\hat{\theta})$ and finally compute the entropy 
$$S(X_i) = -\int_0^\infty G(\hat{\theta}; X_i) \ln{G(\hat{\theta}; X_i)} \mathrm{d}\hat{\theta}.$$
We use windows of $120\times 120$ pixels as potential centers in the map, and use $r_{\text{min}}=30\%$ of the masking radius and $r_{\text{max}}=90\%$ of the masking radius.
We overlay the resulting entropy maps over the patterns in the supplementary figure S1, where our selected center is the point that minimize the entropy, which is marked by the green area.

\section{Aggregation radius}
The most immediate and stark effect of adding salt into the droplets is in the last stage of the drying, where the salted droplets deposit a compact disk instead of continuing the instability.
This transition happens fairly abruptly at all directions (\emph{i.e.} with a circular morphology), at the near-uniform radius we define as the aggregation radius $r_a$.  
We measure these radii to lie between the range $r_a = [0.45, 1.05]$\,mm averaging at $r_a= 0.74\pm0.14$\,mm. 

\begin{figure}
\centering
\includegraphics[width=0.95\linewidth]{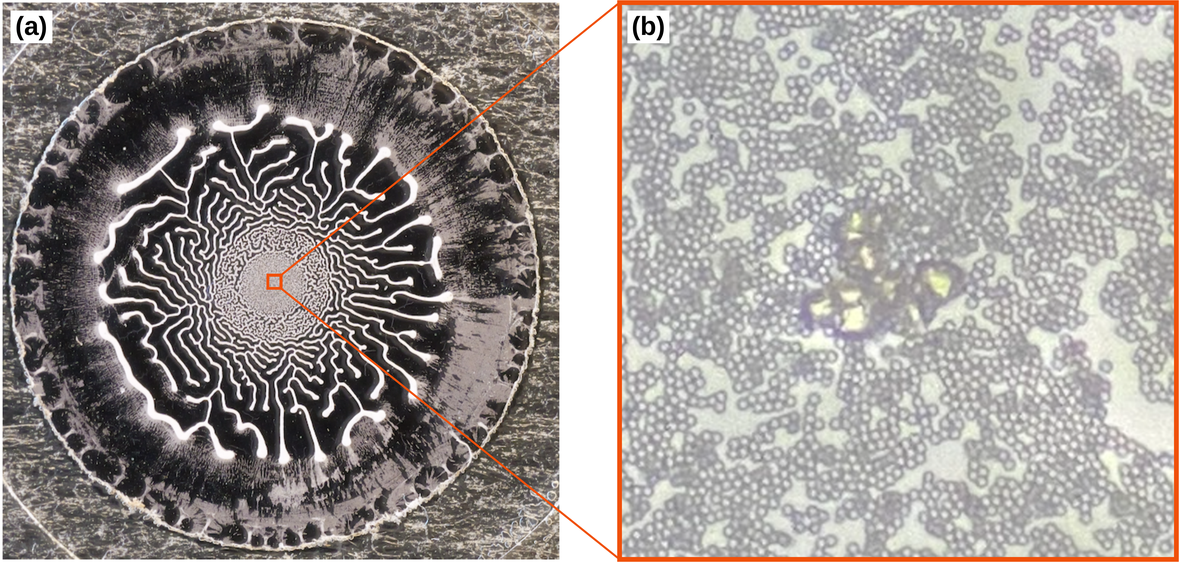}
\caption{Image of a salt crystal in the center of the aggregate in pattern CS4.}
\label{fig:Salt_crystal}
\end{figure}

We hypothesize that these compact colloidal aggregates at the center of the dried CS1-CS12 samples to be mostly caused by the salt, which cannot evaporate. Hence, as the droplet volume shrinks with time $V(t)$, the local salt concentration increase as $C(t)=C_0\big(V_0/V(t)\big)$, thereby increasing the electrostatic screening effect in the assumed DLVO forces between colloids.
The central aggregate is then formed when the DLVO energy barrier becomes comparable with $k_BT$, leaving the colloids stick to each other and forming a (mostly) connected aggregate.
Assuming the droplet to be cylindrical with a fixed height at all times, we may estimate the salt concentration as $C(r)=C_0 \big(r_0/r\big)^2$.
Thus, we estimate the salt concentration at this radius to be $C_a\approx 3$\,M.
Finally, the rest of the liquid evaporates and leave precipitated salt crystals in the central aggregate.
One such crystal is shown in Figure \ref{fig:Salt_crystal}.

\section{Network properties of the fully dried patterns}
Starting from the pioneering works \cite{Murray1926a, Murray1926b, Horton1945, Strahler1957}, the tree-like systems in nature are characterized by the diameters and branching angles at the bifurcations, ordering, the diameter-length and length-area relationships. The network statistics in these systems are optimal when they follow Murray’s law\cite{Murray1926a}, which relates the width of the primary finger to the sum of the width of the fingers after bifurcation as 

\begin{equation}
\label{eq1}
d_{1}^3 +  d_{2}^3  = d_{0}^3 ,
\end{equation}

\noindent and for branching angles in the bifurcation \cite{Murray1926b}

\begin{equation}
\label{eq2}
\cos(\alpha_{1,2}) = \frac{ d_{0}^4 + d_{1,2}^4 - d_{2,1}^4 }{2 d_{0}^2d_{1,2}^2 }.
\end{equation}

\begin{figure}
\centering
\includegraphics[width=0.95\linewidth]{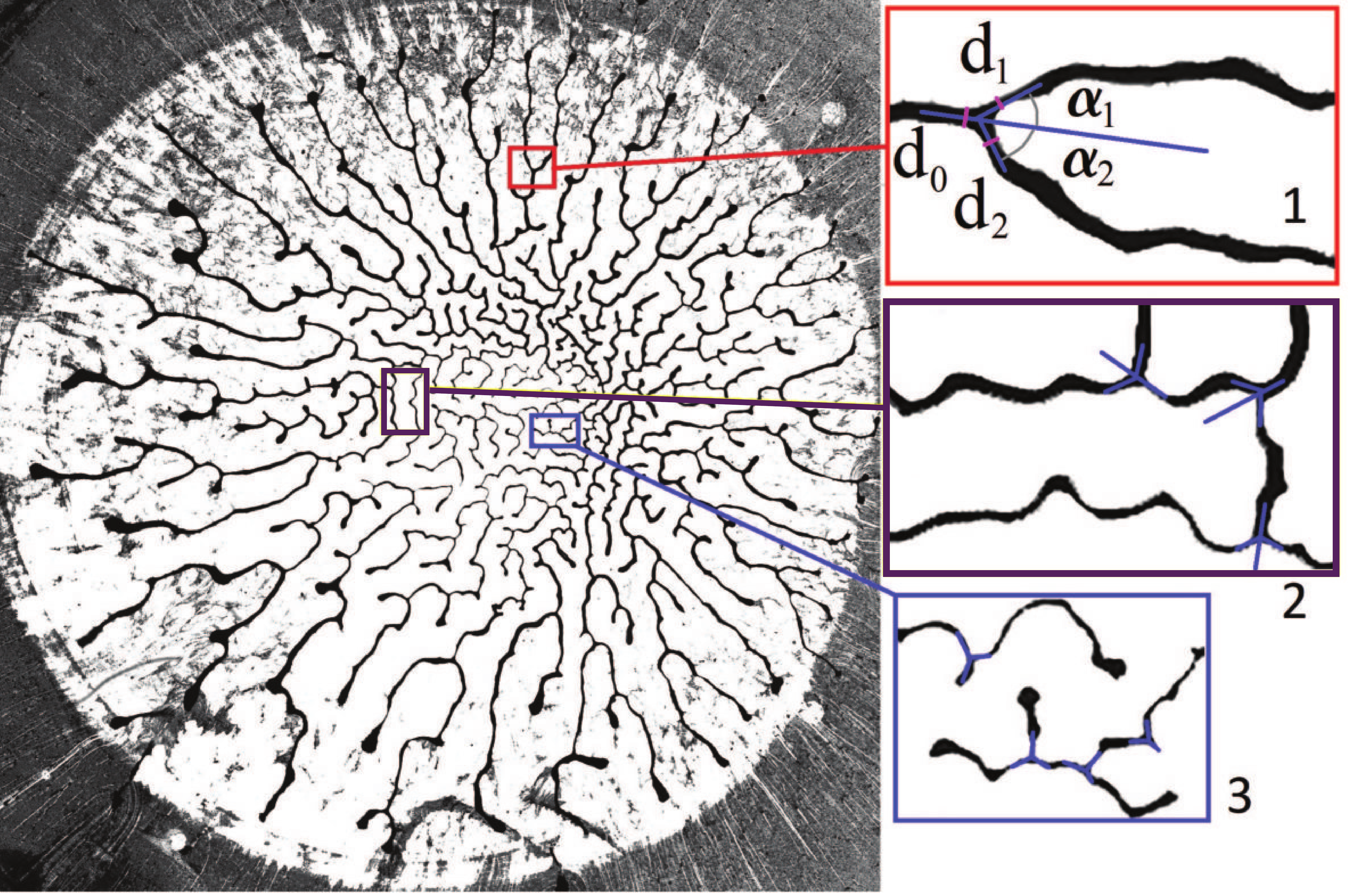}
\caption{Three statistically different solid fingers in an unsalted pattern: radial bifurcations (red square), transversal (purple square) and mixed (blue square).}
\label{fig:Fig_N2}
\end{figure}

\noindent Here $d_{0}$, $d_{1}$, $d_{2}$ are the diameters of the parent and two daughter tubes, $\alpha_{1,2}$ are the branching angles (see Figure \ref{fig:Fig_N2}). For example, if $d_{0} \approx d_{1} \approx d_{2}$ equation \eqref{eq2} gives $\alpha_{1}  \approx \alpha_{2} \approx 60^{\circ}$, and the fingers form a hexagonal-type connection at the angles $120^{\circ}$, which is how 2D foam networks coarsen \cite{weaire1999physics}.

The relationships \eqref{eq1}, \eqref{eq2} are solutions of the optimization problem with minimal total work at a given volume of the structure \cite{Murray1926a, Murray1926b, LaBarbera1990}, but the same relationships hold for the fluid flows in the tubes with porous walls \cite{Kizilova2004} and bifurcations of solid beams with circular cross sections (tree branches) \cite{Kizilova2015}.

\begin{figure}[ht]
\centering
\includegraphics[width=1.0\linewidth]{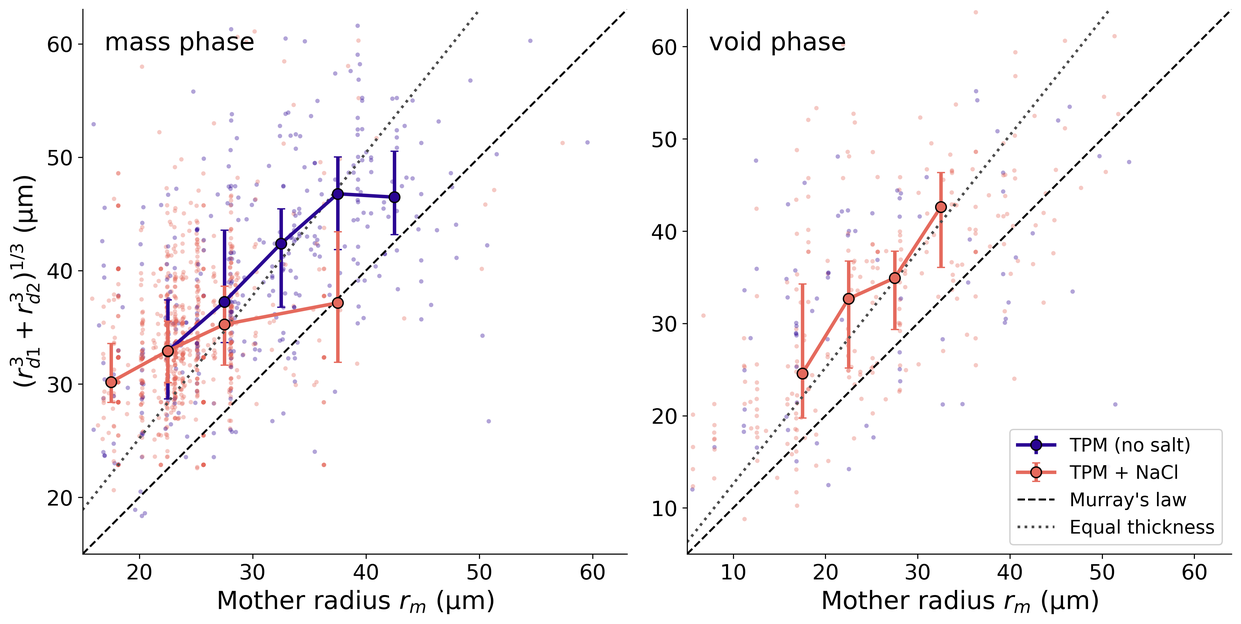}
\caption{Scatter plot of the ratios of all bifurcations in all patterns. The black dashed lines indicate Murray's law \eqref{eq1} is satisfied in the bifurcations, while the dotted line indicate equal thicknesses in all 3 fingers.}
\label{fig:Murray_scatter}
\end{figure}

The shape of the fingers is also different in the salted samples. Qualitatively, we detect three characteristic types of fingers, as seen in Figure \ref{fig:Fig_N2}: Y-shaped tip-junction bifurcations (1), an intermediate region with thinner transversal fingers (2), and a central region with only occasionally oriented tiny fingers (3); all of them form bifurcations. 
The paths along each finger from the region 3 to the center are qualitatively similar to the river systems with numerous tributaries\cite{Devauchelle_etal_2012}.
Therefore, as in the river networks, fingers formed by a conjugation of two outer (daughter) fingers, are regarded as a parent finger, in the meaning of the Murray’s law (1), (2). 
We therefore use the streaming numbers of the branches to determine the mother branch in a bifurcation. 
If there is a tie in streaming numbers, the branch closer to the center is chosen as the mother.

With the skeletonization, we obtain data on the finger thicknesses and branching angles at each bifurcation point, which are plotted in Figure \ref{fig:Murray_scatter}.
We do this by first finding all degree-3 nodes in the skeleton, then traverse a fixed distance from the bifurcation point to measure the thicknesses and angles as shown in Figure \ref{fig:Fig_N2}.
We take our measurements 70\,$\mu$m from each bifurcation for all patterns.
If the branches are shorter than this distance, they are simply disregarded.

\begin{figure}
\centering
\includegraphics[width=1.0\linewidth]{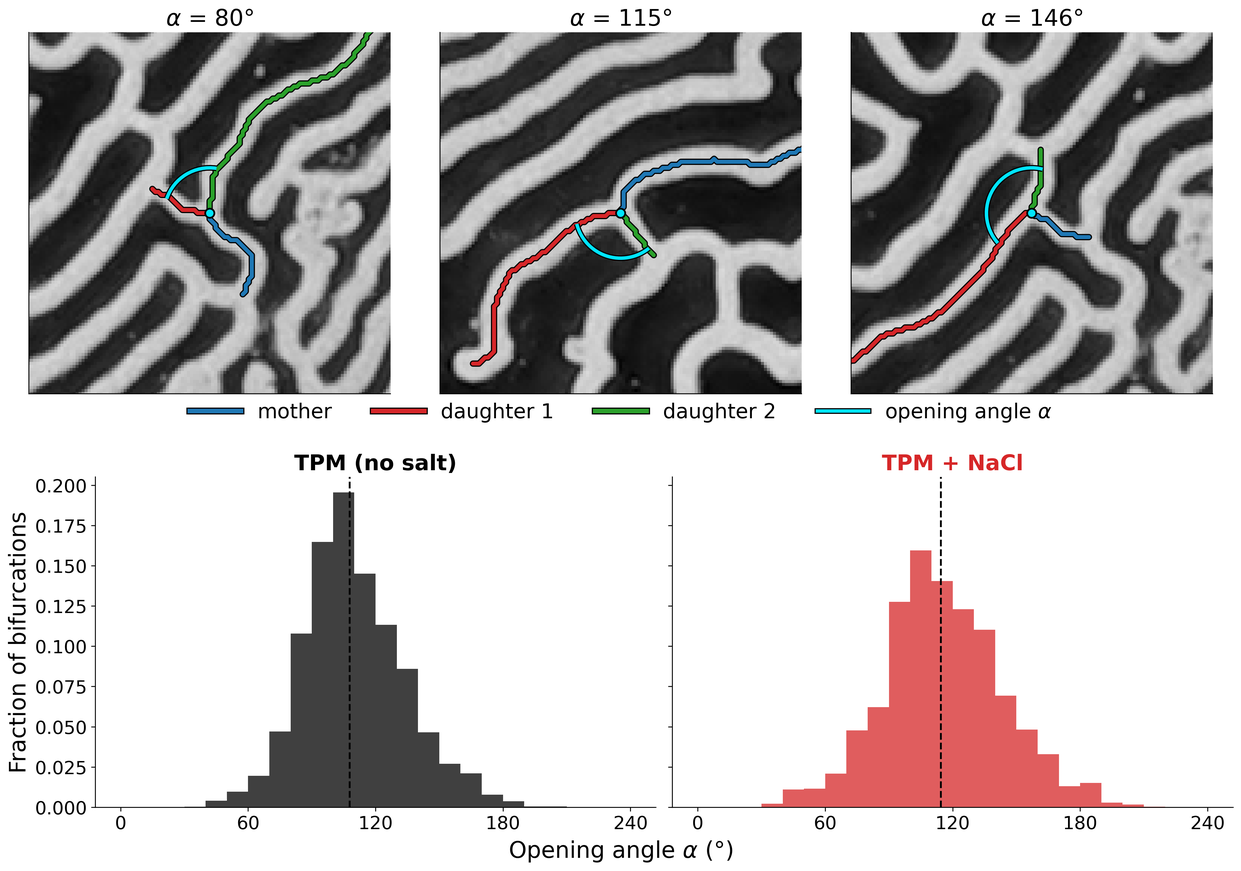}
\caption{Histograms of all total branching angles $\alpha = \alpha_1 + \alpha_2$, in the mass phase (left) and void phase (right), split into unsalted and salted patterns (gray and red, respectively).} 
\label{fig:Angle_distributions}
\end{figure}

Our result show indicate that this network does not follow Murray's law for finger thicknesses, with most fingers growing thinner when branched together.
This is consistent for both salted and unsalted patterns, and follow the general trend that fingers grow thinner towards the center, as found by Henning \textit{et al.}\cite{Hennig_etal_2026a}.
Interestingly, the branching angles do not follow any clear trend, peaking around $\alpha_1 \approx \alpha_2 \approx 54^\circ$, or a total branching angle of $\alpha\approx 108^\circ$ for unsalted patterns and a slightly higher $\alpha_1 \approx \alpha_2 \approx 58^\circ$ with salt.
We show the distributions of branching angles (combined $\alpha = \alpha_1 + \alpha_2$) in \ref{fig:Angle_distributions}.
For comparison Lu et \emph{al.} performed a similar analysis on the nanodroplet networks that emerge from confining the Ouzo effect, which gave a total branching angle of $\alpha \approx 74^\circ$, independent on various control variables\cite{Lu_etal_2017}. 
However, in our case the underlying driving force for such a branching behavior is not yet clear.

\section{Vector-field analysis of finger orientation.}
The solid fingers form by an evaporating interfaces, and are generally directed parallel to the drying direction.
Yet one feature that clearly distinguishes our colloidal patterns from other finger growth phenomena (\emph{e.g.} viscous fingers) is the occurrence of fingers transversal to the general growth direction.

In order to investigate the effect of ions in our droplet, we want to compare the extent of transversal fingers in the screened patterns (\emph{i.e.} salted droplets) to the unscreened case.

We quantify the extent of transversal fingers by defining the directionality $\phi$ of the fingers.
First we split the network into vectors of length $|\vec {v}| = 0.05$\,mm, directed from the outer tips and inwards to the center by using the Horton-Strahler ordering to ensure vectors move towards the (network) center.
The angles of the vectors are compared to the director unit vector $-\hat{r}$ pointing radially inwards to the \emph{entropy center} of the pattern.
We then define the deflection angle as the difference in angle between the vectors following our network and {$\hat{r}$}, such that $\phi = \arccos\big(\big|\frac{\vec{v} \cdot \hat{r}}{|\vec{v}|} \big|\big)$.

\begin{figure*}[ht]
\centering
\includegraphics[width=0.9\linewidth]{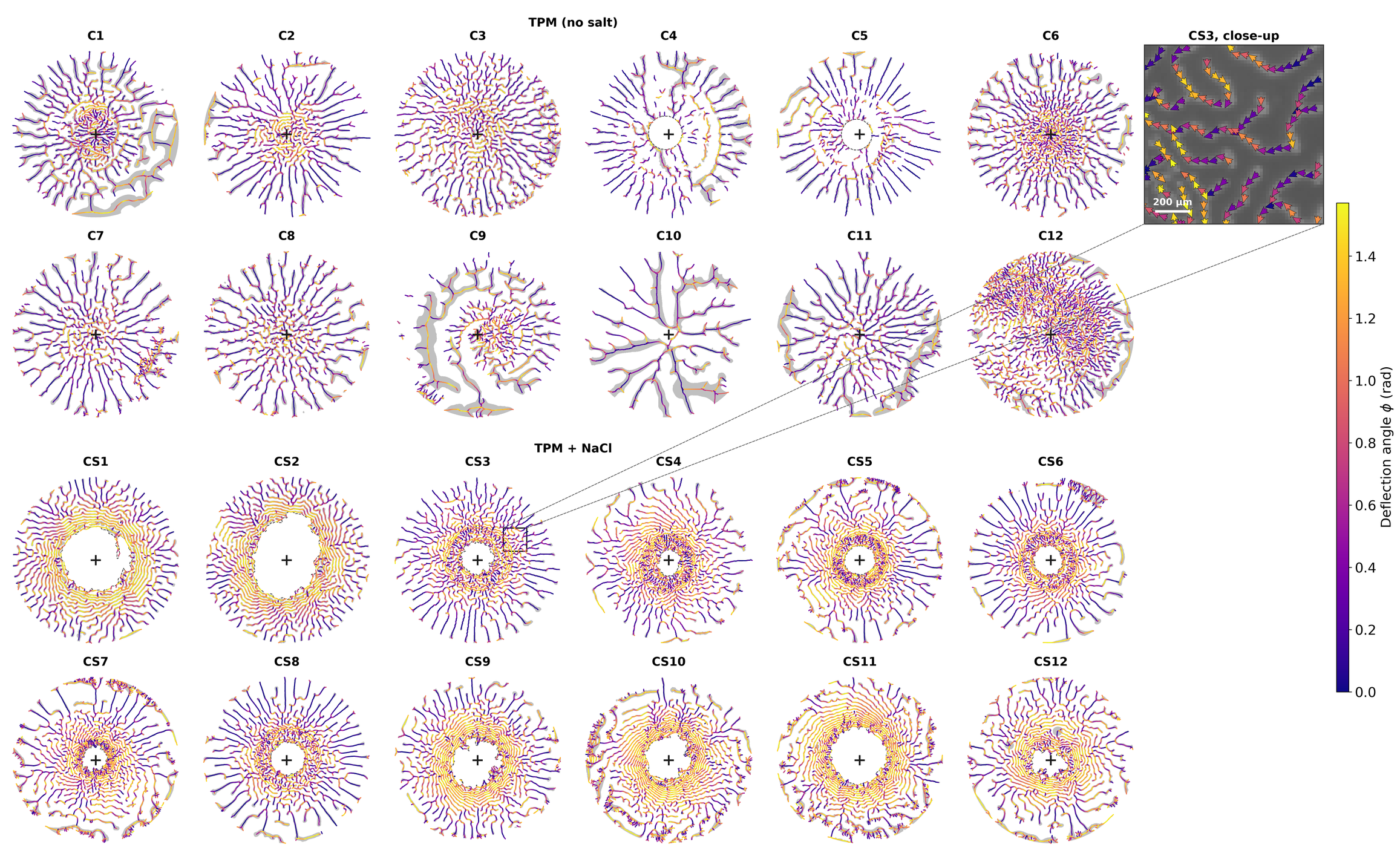}
\caption{Maps of deflection angles $\phi$ throughout the networks of all patterns. The deflection angle is defined as the angle between the vectors of the skeletonized pattern and the vector $-\hat{r}$ pointing towards the center (\emph{i.e.} the drying direction).}
\label{fig:Deflection_angles_all}
\end{figure*}

The deflection angles are shown as a color-coded map above their respective patterns in Figure \ref{fig:Deflection_angles_all}.
As the fingers rarely move \emph{against} the drying direction, the deflection angle is limited to values between 0 (perfectly parallel) and $\pi/2$ (perfectly transversal), by only using the absolute value of the vector product $\hat{r}\cdot\vec{v}$; the few cases of fingers moving away from the center are typically observed in the outer rims of the pattern formation stage and provide no physical insight. 

\begin{table}[ht]
\centering
\begin{tabular}{lcc}
\toprule
Pattern & $\langle\phi\rangle$ (rad) & $\langle \mathrm{Trt}_7 \rangle$ \\
\midrule
C1 & 0.65 & 1.128 \\
C2 & 0.52 & 1.111 \\
C3 & 0.62 & 1.131 \\
C4 & 0.49 & 1.073 \\
C5 & 0.45 & 1.070 \\
C6 & 0.60 & 1.130 \\
C7 & 0.54 & 1.080 \\
C8 & 0.54 & 1.081 \\
C9 & 0.73 & 1.103 \\
C10 & 0.53 & 1.066 \\
C11 & 0.52 & 1.090 \\
C12 & 0.72 & 1.213 \\
CS1 & 0.93 & 1.104 \\
CS2 & 0.90 & 1.128 \\
CS3 & 0.65 & 1.134 \\
CS4 & 0.79 & 1.123 \\
CS5 & 0.82 & 1.112 \\
CS6 & 0.70 & 1.126 \\
CS7 & 0.81 & 1.123 \\
CS8 & 0.65 & 1.115 \\
CS9 & 0.88 & 1.116 \\
CS10 & 0.96 & 1.132 \\
CS11 & 1.00 & 1.126 \\
CS12 & 0.90 & 1.115 \\
\midrule
{\bf Mean unsalted} & 0.58 $\pm$ 0.09 & 1.106 $\pm$ 0.041 \\
{\bf Mean salted} & 0.83 $\pm$ 0.12 & 1.121 $\pm$ 0.009 \\
\bottomrule
\end{tabular}
\caption{Mean deflection angle $\langle\phi\rangle$ and mean local tortuosity $\langle \mathrm{Trt}_7 \rangle$ of the mass phase per pattern. The grouped rows show mean $\pm$ standard deviation over the patterns of the two groups: Unsalted C1-C12 and salted CS1-CS12.}
\label{tab:deflection_trt7_summary}
\end{table}

\begin{figure}
\centering
\includegraphics[width=1.0\linewidth]{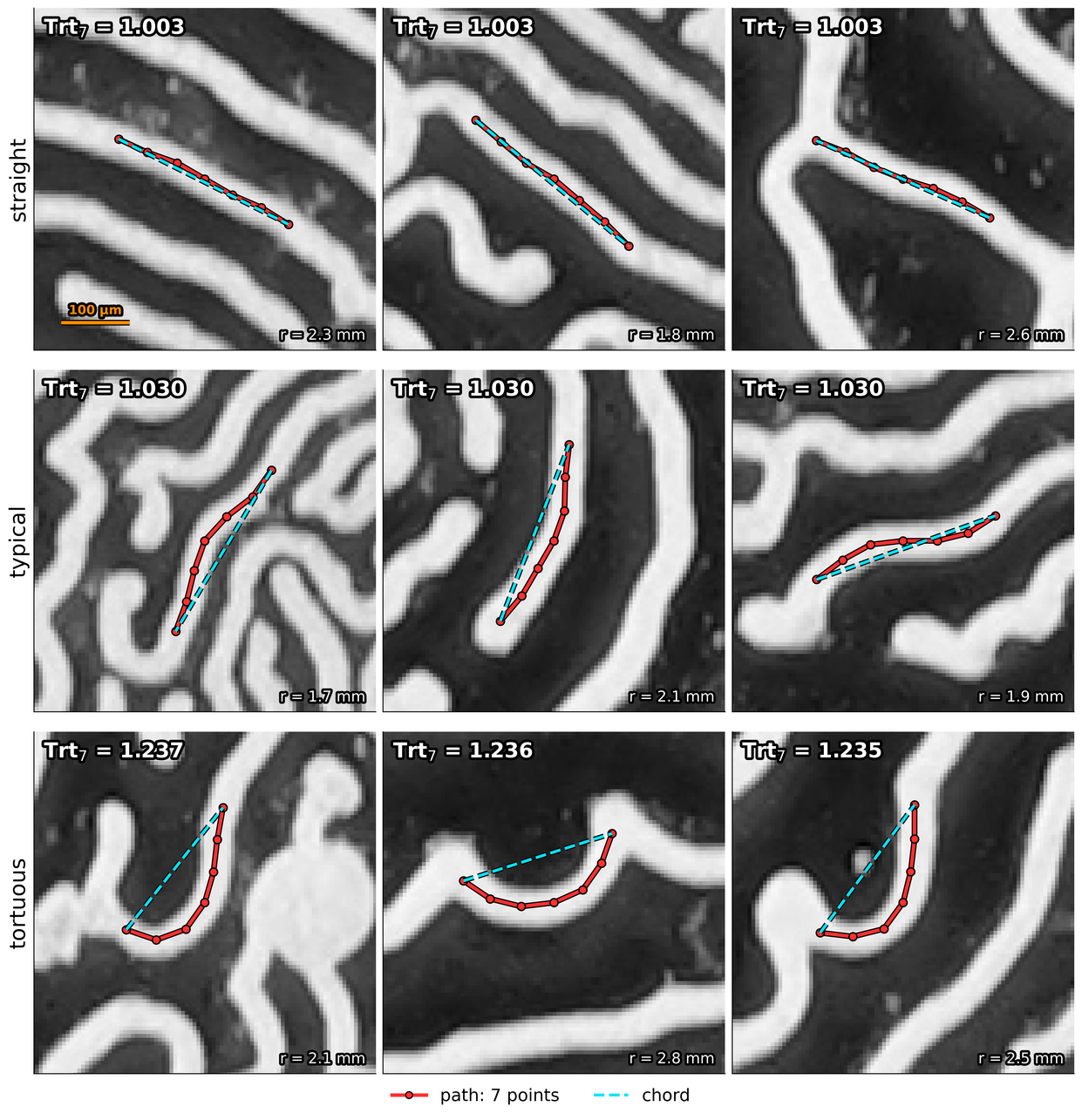}
\caption{Examples of tortuosity calculations with 7 segments; path length is compared with the straight chord. We find no significant difference in local tortuosity when comparing salted and unsalted samples.} 
\label{fig:Tortuosity}
\end{figure}

On average, the screened colloids produced a higher deflection angle than in samples prepared in deionized water.
In particular, these patterns form clear bands of transversal fingers before the fingers merge into one solid central lump.
In the non-saline case, patterns C9 and C12 yield similar \emph{average} deflection angles, but lack the clear band structure.

With these vectors we are also able to probe the local finger tortuosity: the n-point tortuosity is defined as a ratio of the arc length along n vectors to the straight line between these n vectors

$$
Trt_{n}=\frac{1}{|r_{n}-r_{1}|}\sum_{j=1}^{n-1}| r_{j+1}-r_{j}|
$$

For our estimations, $Trt_{7}$ was computed for $n=7$ consecutive points along the solid and void fingers starting from their tips and moving towards the center.
In the cases where we do not have 7 consecutive vectors, the tortuosity is not calculated -- therefore we have a trade-off between the coverage of the tortuosity map and the length we are able to resolve. 
The computed values $Trt_{7}$ was then attributed to the middle point (n=3) of each short chain along the structure, where the mean tortuosity value is presented in Table \ref{tab:deflection_trt7_summary}.
Interestingly, the tortuosity measurements made no meaningful distinction between the compared droplets.
This means that although the overall pattern properties are strongly impacted by the colloidal interactions, the fingers themselves are still fairly similar.

\section{Finger widths compared to the arc-length method}
In a previous work, we proposed using the arc-lengths of these drying patterns to probe the overall statistics\,\cite{Hennig_etal_2026a}. 
However, in the case of the salted patterns, we observe elongated fingers transversal to the drying direction.
These skew the results from the arc-length method, in particular for measuring finger widths, where network methods are more useful.

\begin{figure}
\centering
\includegraphics[width=1.0\linewidth]{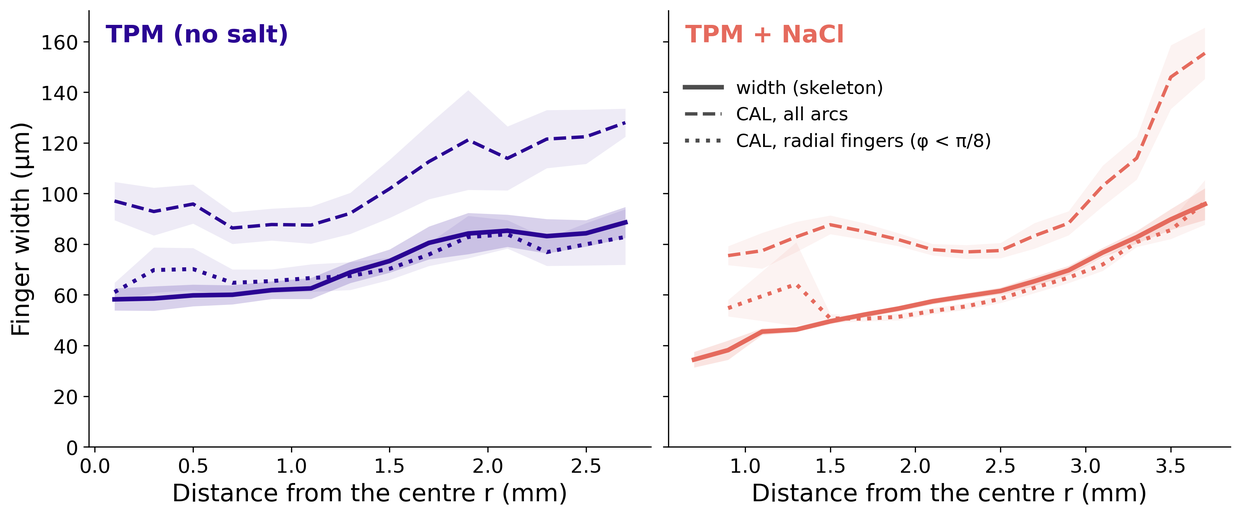}
\caption{Finger widths of the two groups of patterns, plotted against radius. We find that mean arc-lengths overshoot as a measure for finger widths (dashed lines), caused by the transversal fingers, $i.e.$ where the deflection angle $\phi>\pi/8$. We therefore mask these transversal fingers and re-run the CAL method on only the radial fingers (dotted lines), obtaining results closer to the width from the skeleton (solid lines).} 
\label{fig:Finger_Widths}
\end{figure}

We measure the finger thickness over the whole pattern, group them and plot by radius from the center in Figure \ref{fig:Finger_Widths}
The addition of salt make the fingers consistently thinner, with a similar decay as the unsalted ones.

We also observe that the deflection angle may be used as a criterion for the arc-length method.
Figure \ref{fig:Finger_Widths} also show the resulting measurements from using the mean arc-lengths as a measure for finger thicknesses, as a dashed line.
If we omit the transversal fingers by neglecting all points with $\phi \geq \pi/8$, we find a much stronger agreement between skeleton-widths and the mean arc-lengths.

\section{Conclusions and outlook}
Self-assembled colloidal drying patterns offer fascinating labyrinth morphologies in both the colloid phase and the invading air phase. 
As the resultant patterns vary in their shape and morphology, we need to use characterization tools to draw meaningful conclusions. 
In this work, we use skeletonization analysis to study the bifurcations of these patterns and quantify Murray parameters, tortuosity, deflection angle and finger widths. 
We compare two types of patterns: a ``base" case, taken as those studied previously\cite{Beechey-Newman_etal_2025, Hennig_etal_2026a}, and patterns formed with charge-screened colloids.
Upon comparison, these patterns vary on some measures, and are similar on others. 
The opening angles were found to be similar for both groups of patterns ($54^\circ$\, for unsalted; $58^\circ$ for salted), and the finger tortuosity did not change with screening lengths.
On the other hand, the salted droplets are characterized by an abrupt change in the finger formation -- transitioning to a solid center at an aggregation radius $r_a$, which we measured to $0.74\pm0.14$\,mm.

We interpret these findings as an indicator that the instability and resulting patterns not only depend on fluid properties (wetting, viscosity) or system properties (drying rate, humidity), but also depend on colloid-colloid-interactions implying that screened self-assembled patterns are different than unscreened patterns.

\section*{Conflicts of interest}
There are no conflicts to declare.

\section*{Data Availability}
The codes we use for our information entropy maps is available in a Zenodo repository \cite{hennig_2026_18954724}. All other code and experimental images are available upon request.

%
%

\section*{Acknowledgements}
This work was supported by the Research Council of Norway through its Center of Excellence funding scheme, project number RCN 262644 and the INTPART program, Project No. RCN 309139. 
EE was funded from the European Union's Horizon Europe research and innovation programme under the Marie Skłodowska-Curie Doctoral Network FLUXIONIC, grant agreement No. 101119598.



\balance

\bibliography{rsc}

@article{Adachi_etal_1995,
  title = {Stripe {{Patterns Formed}} on a {{Glass Surface}} during {{Droplet Evaporation}}},
  author = {Adachi, Eiki and Dimitrov, Antony S. and Nagayama, Kuniaki},
  year = {1995},
  month = apr,
  journal = {Langmuir},
  volume = {11},
  number = {4},
  pages = {1057--1060},
  publisher = {American Chemical Society},
  issn = {0743-7463},
  doi = {10.1021/la00004a003},
  urldate = {2025-04-28}
}

@article{Majumder_etal_2012,
  title = {Overcoming the ``{{Coffee-Stain}}'' {{Effect}} by {{Compositional Marangoni-Flow-Assisted Drop-Drying}}},
  author = {Majumder, Mainak and Rendall, Clint S. and Eukel, J. Alexander and Wang, James Y. L. and Behabtu, Natnael and Pint, Cary L. and Liu, Tzu-Yu and Orbaek, Alvin W. and Mirri, Francesca and Nam, Jaewook and Barron, Andrew R. and Hauge, Robert H. and Schmidt, Howard K. and Pasquali, Matteo},
  year = {2012},
  month = jun,
  journal = {The Journal of Physical Chemistry B},
  volume = {116},
  number = {22},
  pages = {6536--6542},
  publisher = {American Chemical Society},
  issn = {1520-6106},
  doi = {10.1021/jp3009628},
  urldate = {2024-11-06}
}

@article{Anyfantakis_etal_2015,
  title = {Modulation of the {{Coffee-Ring Effect}} in {{Particle}}/{{Surfactant Mixtures}}: The {{Importance}} of {{Particle}}--{{Interface Interactions}}},
  shorttitle = {Modulation of the {{Coffee-Ring Effect}} in {{Particle}}/{{Surfactant Mixtures}}},
  author = {Anyfantakis, Manos and Geng, Zheng and Morel, Mathieu and Rudiuk, Sergii and Baigl, Damien},
  year = {2015},
  month = apr,
  journal = {Langmuir},
  volume = {31},
  number = {14},
  pages = {4113--4120},
  publisher = {American Chemical Society},
  issn = {0743-7463},
  doi = {10.1021/acs.langmuir.5b00453},
  urldate = {2025-04-28}
}

@article{Ben_Demekhin_2002,
  title = {A Spectral Theory for Small-Amplitude Miscible Fingering},
  author = {Ben, Yuxing and Demekhin, Evgeny},
  year = {2002},
  month = mar,
  journal = {Physics of Fluids},
  volume = {14},
  pages = {999--1010},
  doi = {10.1063/1.1446885}
}

@unpublished{beechey2026elasticity,
  title={Effect of elasticity on the drying pattern of confined colloidal droplets},
  author={Beechey-Newman, Ilaria and Hennig, Andreas Andersen and Dumay, Yann and Tammen, Ren{\'e} and Eiser, Erika},
  note={Manuscript in preparation},
  year={2026}
}

@article{Bigerelle_etal_2006,
  title = {Monte {{Carlo}} Simulation of Gold Nano-Colloids Aggregation Morphologies on a Heterogeneous Surface},
  author = {Bigerelle, M. and Haidara, H. and Van Gorp, A.},
  year = {2006},
  month = jul,
  journal = {Materials Science and Engineering: C},
  series = {Current {{Trends}} in {{Nanoscience}} - from {{Materials}} to {{Applications}}},
  volume = {26},
  number = {5},
  pages = {1111--1116},
  issn = {0928-4931},
  doi = {10.1016/j.msec.2005.09.043},
  urldate = {2025-04-28}
}

@article{Beechey-Newman_etal_2025,
  title = {Confined Colloidal Droplets Dry to Form Circular Mazes},
  author = {Beechey-Newman, Ilaria and Kizilova, Natalya and Hennig, Andreas Andersen and Flekkøy, Eirik Grude and Eiser, Erika},
  date = {2025-08-12},
  year = {2025},
  journal = {Proceedings of the National Academy of Sciences},
  volume = {122},
  number = {32},
  pages = {e2508363122},
  publisher = {Proceedings of the National Academy of Sciences},
  doi = {10.1073/pnas.2508363122},
  url = {https://www.pnas.org/doi/full/10.1073/pnas.2508363122},
  urldate = {2025-08-11}
}

@article{Brust_etal_1998,
  title = {Self-{{Assembled Gold Nanoparticle Thin Films}} with {{Nonmetallic Optical}} and {{Electronic Properties}}},
  author = {Brust, Mathias and Bethell, Donald and Kiely, Christopher J. and Schiffrin, David J.},
  year = {1998},
  month = sep,
  journal = {Langmuir},
  volume = {14},
  number = {19},
  pages = {5425--5429},
  publisher = {American Chemical Society},
  issn = {0743-7463},
  doi = {10.1021/la980557g},
  urldate = {2025-04-28}
}

@article{Chen_etal_2005a,
  title = {Fabrication and Luminescence of {{ZnS}} : {{Mn2}}+ Nanoflowers},
  author = {Chen, W. and Bovin, J.-O. and Wang, S. P. and Joly, A. G. and Wang, Y. Q. and Sherwood, P. M. A.},
  year = {2005},
  journal = {Journal of Nanoscience and Nanotechnology},
  volume = {5},
  number = {9},
  pages = {1309--1322},
  doi = {10.1166/jnn.2005.302}
}

@article{Chen_etal_2016,
  title = {Blood Drop Patterns: {{Formation}} and Applications},
  author = {Chen, R. and Zhang, L. and Zang, D. and Shen, W.},
  year = {2016},
  journal = {Advances in Colloid and Interface Science},
  volume = {231},
  pages = {1--14},
  doi = {10.1016/j.cis.2016.01.008}
}

@article{Deegan_2000a,
  title = {Pattern Formation in Drying Drops},
  author = {Deegan, Robert D.},
  year = {2000},
  journal = {Physical Review E},
  volume = {61},
  pages = {475--85}
}

@article{Deegan_etal_1997a,
  title = {Capillary Flow as the Cause of Ring Stains from Dried Liquid Drops},
  author = {Deegan, Robert D and Bakajin, Olgica and Dupont, Todd F and Huber, Greb and Nagel, Sidney R and Witten, Thomas A},
  year = {1997},
  journal = {Nature},
  volume = {389},
  number = {6653},
  pages = {827--829},
  publisher = {Nature Publishing Group UK London}
}

@article{Dudala_Goel_2024,
  title = {Droplet and Digital Microfluidics},
  author = {Dudala, S. and Goel, S.},
  year = {2024},
  journal = {Ideation to Implementation. - Digital microfluidics},
  pages = {49--76},
  doi = {10.1016/B978-0-443-15416-4.00005-0}
}

@article{Fang_etal_2002,
  title = {Complex Pattern Formation by Cowpea Mosaic Virus Nanoparticles},
  author = {Fang, J. and Soto, C. M. and Lin, T. and Johnson, J. E. and Ratna, B.},
  year = {2002},
  journal = {Langmuir : the ACS journal of surfaces and colloids},
  volume = {18},
  number = {2},
  pages = {308--310},
  doi = {10.1021/la0112357}
}

@article{Fang_etal_2006,
  title = {Drying of {{DNA}} Droplets},
  author = {Fang, X. and Li, B. and Petersen, E. and Seo, Y.-S. and Samuilov, V. A. and Chen, Y. and others and Rafailovich, M. H.},
  year = {2006},
  journal = {Langmuir : the ACS journal of surfaces and colloids},
  volume = {22},
  number = {14},
  pages = {6308--6312},
  doi = {10.1021/la060479u}
}

@article{Haidara_etal_2001,
  title = {Spontaneous Growth of Two-Dimensional Complex Patterns of Nanoparticles at Model Molecular Surfaces},
  author = {Haidara, H. and Mougin, K. and Schultz, J.},
  year = {2001},
  journal = {Langmuir : the ACS journal of surfaces and colloids},
  volume = {17},
  number = {3},
  pages = {659--663},
  doi = {10.1021/la0012034}
}

@article{Harris_etal_2007,
  title = {Patterning Colloidal Films via Evaporative Lithography},
  author = {Harris, D. J. and Hu, H. and Conrad, J. C. and Lewis, J. A.},
  year = {2007},
  journal = {Physical Review Letters},
  volume = {98},
  number = {14},
  pages = {148301},
  doi = {10.1103/physrevlett.98.148301}
}

@article{Hu_Larson_2006a,
  title = {Marangoni Effect Reverses Coffee-Ring Depositions},
  author = {Hu, H. and Larson, R. G.},
  year = {2006},
  journal = {The Journal of Physical Chemistry B},
  volume = {110},
  number = {14},
  pages = {7090--7094},
  doi = {10.1021/jp0609232}
}

@article{Jaber_etal_2021,
  title = {Effect of Temperature on Evaporation Dynamics of Sheep's Blood Droplets and Topographic Analysis of Induced Patterns},
  author = {Jaber, A. and Vayron, R. and Harmand, S.},
  year = {2021},
  journal = {Heliyon},
  volume = {8},
  number = {11},
  pages = {e11258}
}

@article{Jang_Oh_2004,
  title = {Morphogenesis of Evaporation-Induced Self-Assemblies of Polypyrrole Nanoparticles Dispersed in a Liquid Medium},
  author = {Jang, J. and Oh, J. H.},
  year = {2004},
  journal = {Langmuir : the ACS journal of surfaces and colloids},
  volume = {20},
  number = {20},
  pages = {8419--8422},
  doi = {10.1021/la049259b}
}

@article{Khatri_etal_2008,
  title = {Structural Organization of Gold Nanoparticles onto the {{ITO}} Surface and Its Optical Properties as a Function of Ensemble Size},
  author = {Khatri, O. P. and Murase, K. and Sugimura, H.},
  year = {2008},
  journal = {Langmuir : the ACS journal of surfaces and colloids},
  volume = {24},
  number = {8},
  pages = {3787--3793},
  doi = {10.1021/la7039042}
}

@article{Kim_etal_2016,
  title = {Controlled Uniform Coating from the Interplay of Marangoni Flows and Surface-Adsorbed Macromolecules},
  author = {Kim, H. and Boulogne, F. and Um, E. and Jacobi, I. and Button, E. and Stone, H. A.},
  year = {2016},
  journal = {Physical Review Letters},
  volume = {116},
  number = {12},
  pages = {124501},
  doi = {10.1103/physrevlett.116.124501}
}

@article{Lim_etal_2010,
  title = {Nucleation and Growth Mechanisms for {{Pd-Pt}} Bimetallic Nanodendrites and Their Electrocatalytic Properties},
  author = {Lim, B. and Jiang, M. and Yu, . T. and Camargo, P. H. C. and Xia, Y.},
  year = {2010},
  journal = {Nano Research},
  volume = {3},
  number = {2},
  pages = {69--80},
  doi = {10.1007/s12274-010-1010-8}
}

@article{Maeda_1999,
  title = {An Atomic Force Microscopy Study of Ordered Molecular Assemblies and Concentric Ring Patterns from Evaporating Droplets of Collagen Solutions},
  author = {Maeda, H.},
  year = {1999},
  journal = {Langmuir : the ACS journal of surfaces and colloids},
  volume = {15},
  number = {24},
  pages = {8505--8513},
  doi = {10.1021/la981738l}
}

@article{Maeda_2000,
  title = {Observation of Spatially Rhythmic Patterns from Evaporating Collagen Solution Droplets},
  author = {Maeda, H.},
  year = {2000},
  journal = {Langmuir : the ACS journal of surfaces and colloids},
  volume = {16},
  number = {26},
  pages = {9977--9982},
  doi = {10.1021/la0001326}
}

@article{Maheshwari_etal_2008,
  title = {Coupling between Precipitation and Contact-Line Dynamics: {{Multiring}} Stains and Stick-Slip Motion},
  author = {Maheshwari, S. and Zhang, L. and Zhu, Y. and Chang, H.-C.},
  year = {2008},
  journal = {Physical Review Letters},
  volume = {100},
  number = {4},
  pages = {044503},
  doi = {10.1103/physrevlett.100.044503}
}

@article{Maillard_etal_2001,
  title = {Rings and Hexagons Made of Nanocrystals},
  author = {Maillard, M. and Motte, L. and Pileni, M.-P.},
  year = {2001},
  journal = {Advanced Materials},
  volume = {13},
  number = {3},
  pages = {200--204},
  doi = {10.1002/1521-4095(200102)13:33.0.co;2-p}
}

@article{Mampallil_Eral_2018,
  title = {A Review on Suppression and Utilization of the Coffee-Ring Effect},
  author = {Mampallil, D. and Eral, H. B.},
  year = {2018},
  journal = {Advances in Colloid and Interface Science},
  volume = {252},
  pages = {38--54},
  doi = {10.1016/j.cis.2017.12.008}
}

@article{Mougin_Haidara_2002,
  title = {Complex Pattern Formation in Drying Dispersions},
  author = {Mougin, K. and Haidara, H.},
  year = {2002},
  journal = {Langmuir : the ACS journal of surfaces and colloids},
  volume = {18},
  number = {24},
  pages = {9566--9569},
  doi = {10.1021/la020491w}
}

@article{Nguyen_Stebe_2002,
  title = {Patterning of Small Particles by a Surfactant-Enhanced Marangoni-B{\'e}nard Instability},
  author = {Nguyen, V. X. and Stebe, K. J.},
  year = {2002},
  journal = {Physical Review Letters},
  volume = {88},
  number = {16},
  pages = {164501},
  doi = {10.1103/physrevlett.88.164501}
}

@article{Pal_etal_2020,
  title = {Concentration-Driven Phase Transition and Self-Assembly in Drying Droplets of Diluting Whole Blood},
  author = {Pal, Anusuya and Gope, Amalesh and Obayemi, John D and Iannacchione, Germano S},
  year = {2020},
  journal = {Scientific reports},
  volume = {10},
  number = {1},
  pages = {18908},
  publisher = {Nature Publishing Group UK London}
}

@article{Park_Moon_2006,
  title = {Control of Colloidal Particle Deposit Patterns within Picoliter Droplets Ejected by Ink-Jet Printing},
  author = {Park, J. and Moon, J.},
  year = {2006},
  journal = {Langmuir : the ACS journal of surfaces and colloids},
  volume = {22},
  number = {8},
  pages = {3506--3513},
  doi = {10.1021/la053450j}
}

@article{Poulard_etal_2003,
  title = {Freely Receding Evaporating Droplets},
  author = {Poulard, C. and B{\'e}nichou, O. and Cazabat, A. M.},
  year = {2003},
  journal = {Langmuir : the ACS journal of surfaces and colloids},
  volume = {19},
  number = {21},
  pages = {8828--8834},
  doi = {10.1021/la030162j}
}

@article{Rabani_etal_2003,
  title = {Drying-Mediated Self-Assembly of Nanoparticles},
  author = {Rabani, E. and Reichman, D. R. and Geissler, P. L. and Brus, L. E.},
  year = {2003},
  journal = {Nature},
  volume = {426},
  pages = {271--274},
  doi = {10.1038/nature02087}
}

@article{Sefiane_etal_2021,
  title = {Patterns from Dried Drops as a Characterisation and Healthcare Diagnosis Technique, Potential and Challenges: {{A}} Review},
  author = {Sefiane, K. and Duursma, G. and Arif, A.},
  year = {2021},
  journal = {Advances in Colloid and Interface Science},
  volume = {298},
  pages = {102546},
  doi = {10.1016/j.cis.2021.102546}
}

@article{Seknazi_etal_2019,
  title = {From Spinodal Decomposition to Alternating Layered Structure within Single Crystals of Biogenic Magnesium Calcite},
  author = {Seknazi, E. and Kozachkevich, S. and Polishchuk, I. and Bianco Stein, N. and Villanova, J. and Suuronen, J.-P. and others and Pokroy, B.},
  year = {2019},
  journal = {Nature Communications},
  volume = {10},
  number = {1},
  pages = {4559},
  doi = {10.1038/s41467-019-12168-8}
}

@article{Seo_etal_2017,
  title = {Altering the Coffee-Ring Effect by Adding a Surfactant-like Viscous Polymer Solution},
  author = {Seo, C. and Jang, D. and Chae, J. and Shin, S.},
  year = {2017},
  journal = {Scientific Reports},
  volume = {7},
  number = {1},
  pages = {500},
  doi = {10.1038/s41598-017-00497-x}
}

@article{Shen_etal_2010,
  title = {Minimal Size of Coffee Ring Structure},
  author = {Shen, X. and Ho, C.-M. and Wong, T.-S.},
  year = {2010},
  journal = {The Journal of Physical Chemistry B},
  volume = {114},
  number = {16},
  pages = {5269--5274},
  doi = {10.1021/jp912190v}
}

@article{Shmuylovich_etal_2002,
  title = {Surface Morphology of Drying Latex Films: Multiple Ring Formation},
  author = {Shmuylovich, L. and Shen, A. Q. and Stone, H. A.},
  year = {2002},
  journal = {Langmuir : the ACS journal of surfaces and colloids},
  volume = {18},
  number = {9},
  pages = {3441--3445},
  doi = {10.1021/la011484v}
}

@article{Smalyukh_etal_2006,
  title = {Structure and Dynamics of Liquid Crystalline Pattern Formation in Drying Droplets of {{DNA}}},
  author = {Smalyukh, I. I. and Zribi, O. V. and Butler, J. C. and Lavrentovich, O. D. and Wong, G. C. L.},
  year = {2006},
  journal = {Physical Review Letters},
  volume = {96},
  number = {17},
  pages = {177801},
  doi = {10.1103/physrevlett.96.177801}
}

@article{Sommer_etal_2004,
  title = {Limits of the Impact of Gravity on Self-Organizing Nanospheres},
  author = {Sommer, A. P. and {Ben-Moshe}, M. and Magdassi, S.},
  year = {2004},
  journal = {Journal of Physical Chemistry B},
  volume = {108},
  pages = {8--10},
  doi = {10.1021/jp0487049}
}

@article{Sultan_etal_2005,
  title = {Evaporation of a Thin Film: Diffusion of the Vapour and {{Marangoni}} Instabilities},
  author = {Sultan, E. and Boudaoud, A. and Ben Amar, M.},
  year = {2005},
  journal = {Journal of Fluid Mechanics},
  volume = {543},
  pages = {183--202},
  doi = {10.1017/S0022112005006348}
}

@article{Takhistov_Chang_2002,
  title = {Complex Stain Morphologies},
  author = {Takhistov, P. and Chang, H.-C.},
  year = {2002},
  journal = {Industrial \& Engineering Chemistry Research},
  volume = {41},
  number = {25},
  pages = {6256--6269},
  doi = {10.1021/ie010788+}
}

@article{Tang_etal_2002,
  title = {{{Gas}}-liquid-solid Phase Transition Model for Two-Dimensional Nanocrystal Self-Assembly on Graphite},
  author = {Tang, J. and Ge, G. and Brus, L. E.},
  year = {2002},
  journal = {The Journal of Physical Chemistry B},
  volume = {106},
  number = {22},
  pages = {5653--5658},
  doi = {10.1021/jp014559c}
}

@article{Thampi_Basavaraj_2023,
  title = {Drying Drops of Colloidal Dispersions},
  author = {Thampi, Sumesh P and Basavaraj, Madivala G},
  year = {2023},
  journal = {Annual Review of Chemical and Biomolecular Engineering},
  volume = {14},
  number = {1},
  pages = {53--83},
  publisher = {Annual Reviews}
}

@article{VanDerWel_etal_2017,
  title = {Preparation of Colloidal Organosilica Spheres through Spontaneous Emulsification},
  author = {Van Der Wel, Casper and Bhan, Rohit K and Verweij, Ruben W and Frijters, Hans C and Gong, Zhe and Hollingsworth, Andrew D and Sacanna, Stefano and Kraft, Daniela J},
  year = {2017},
  journal = {Langmuir : the ACS journal of surfaces and colloids},
  volume = {33},
  number = {33},
  pages = {8174--8180},
  publisher = {ACS Publications}
}

@article{Wang_etal_2001,
  title = {Surface Patterned Porous Films by Convection-Assisted Dynamic Self-Assembly of Zeolite Nanoparticles},
  author = {Wang, H. and Wang, Z. and Huang, L. and Mitra, A. and Yan, Y.},
  year = {2001},
  journal = {Langmuir : the ACS journal of surfaces and colloids},
  volume = {17},
  number = {9},
  pages = {2572--2574},
  doi = {10.1021/la0102509}
}

@book{Wang_etal_2018,
  title = {Advanced Colloidal Lithography: {{From}} Patterning to Applications},
  author = {Wang, Y. and Zhang, M. and Lai, Y. and Chi, L.},
  year = {2018},
  publisher = {Nano Today},
  doi = {10.1016/j.nantod.2018.08.010}
}

@article{Yang_etal_2014,
  title = {From Multi-Ring to Spider Web and Radial Spoke: Competition between the Receding Contact Line and Particle Deposition in a Drying Colloidal Drop},
  author = {Yang, X. and Li, C. Y. and Sun, Y.},
  year = {2014},
  journal = {Soft Matter},
  volume = {10},
  number = {25},
  pages = {4458--4463},
  doi = {10.1039/c4sm00497c}
}

@article{Yang_etal_2015,
  title = {Shear Banding in Drying Films of Colloidal Nanoparticles},
  author = {Yang, B. and Sharp, J. S. and Smith, M. I.},
  year = {2015},
  journal = {ACS Nano},
  volume = {9},
  number = {4},
  pages = {4077--4084},
  doi = {10.1021/acsnano.5b00127}
}

@article{Yao_etal_2023,
  title = {General Behavior and Mechanisms of Dendritic Crystal Formation in Evaporated Biological Solutions and Their Significance for Solution Analysis},
  author = {Yao, X. and Lei, J.-M. and Huang, Y.-X.},
  year = {2023},
  journal = {Journal of Materials Research and Technology},
  volume = {24},
  pages = {4024--4036},
  doi = {10.1016/j.jmrt.2023.04.045}
}

@article{Zhang_etal_2008,
  title = {Evaporative Self-Assembly from Complex {{DNA}}-colloid Suspensions},
  author = {Zhang, L. and Maheshwari, S. and Chang, H.-C. and Zhu, Y.},
  year = {2008},
  journal = {Langmuir : the ACS journal of surfaces and colloids},
  volume = {24},
  number = {8},
  pages = {3911--3917},
  doi = {10.1021/la703539k}
}

@article{Zhang_etal_2016,
  title = {Forming Nanoparticle Monolayers at Liquid--Air Interfaces by Using Miscible Liquids},
  author = {Zhang, D. and Hu, J. and Kennedy, K. M. and Herman, I. P.},
  year = {2016},
  journal = {Langmuir : the ACS journal of surfaces and colloids},
  volume = {32},
  number = {33},
  pages = {8467--8472},
  doi = {10.1021/acs.langmuir.6b00828}
}

@article{Zhong_Duan_2015,
  title = {Surfactant-Adsorption-Induced Initial Depinning Behavior in Evaporating Water and Nanofluid Sessile Droplets},
  author = {Zhong, X. and Duan, F.},
  year = {2015},
  journal = {Langmuir : the ACS journal of surfaces and colloids},
  volume = {31},
  number = {19},
  pages = {5291--5298},
  doi = {10.1021/acs.langmuir.5b00288}
}

@article{Otsu_1979,
  title = {A {{Threshold Selection Method}} from {{Gray-Level Histograms}}},
  author = {Otsu, Nobuyuki},
  year = {1979},
  month = jan,
  journal = {IEEE Transactions on Systems, Man, and Cybernetics},
  volume = {9},
  number = {1},
  pages = {62--66},
  issn = {2168-2909},
  doi = {10.1109/TSMC.1979.4310076},
  urldate = {2024-12-16}
}

@article{Hoshen_Kopelman_1976,
  title = {Percolation and Cluster Distribution. {{I}}. {{Cluster}} Multiple Labeling Technique and Critical Concentration Algorithm},
  author = {Hoshen, J. and Kopelman, R.},
  year = {1976},
  month = oct,
  journal = {Physical Review B},
  volume = {14},
  number = {8},
  pages = {3438--3445},
  issn = {0556-2805},
  doi = {10.1103/PhysRevB.14.3438},
  urldate = {2024-04-09},
  copyright = {http://link.aps.org/licenses/aps-default-license},
  langid = {english}
}

@article{Eloy2017,
  author = {Eloy, C. and Fournier, M. and Lacointe, A. and Moulia, B.},
  year = {2017},
  title = {Wind loads and competition for light sculpt trees into self-similar structures},
  journal = {Nat Commun},
  volume = {8},
  pages = {1014},
  note = {doi:10.1038/s41467-017-00995-6.},
}

@article{Horsfield1985,
  author = {Horsfield, K.},
  year = {1985},
  title = {Diameters, generations, and orders of branches in the bronchial tree},
  journal = {J Appl Physiol},
  volume = {68},
  number = {2},
  pages = {457-461},
  note = {doi:10.1152/jappl.1990.68.2.457.},
}

@article{Horton1945,
  author = {Horton, R. E.},
  year = {1945},
  title = {Erosional development of streams and their drainage basins: hydrophysical approach to quantitative morphology},
  journal = {GSA Bulletin},
  volume = {56},
  number = {3},
  pages = {275-370},
  note = {doi:10.1130/0016-7606(1945)56[275:EDOSAT]2.0.CO;2.},
}

@book{Kizilova2004,
  author = {Kizilova, N.},
  year = {2004},
  title = {Computational approach to optimal transport network construction in biomechanics},
  series = {Lecture Notes in Computer Science},
  publisher = {--  - Vol. 3044. -- P.476-485},
  note = {doi:10.1007/978-3-540-24709-8\_51.},
}

@article{Kolivand2019,
  author = {Kolivand, H. and Fern, B. M. and Saba, T. and Rahim, M. and Rehman, A.},
  year = {2019},
  title = {A New Leaf Venation Detection Technique for Plant Species Classification},
  journal = {Arab J Sci Eng},
  volume = {44},
  pages = {3315-3327},
  note = {doi:10.1007/s13369-018-3504-8.},
}

@article{Liu2020,
  author = {Liu, H. and Xiang Cao and Xiaohan Yang and Mingxian Liu and Yuri Lvov},
  year = {2020},
  title = {Formation of Regular Wormlike Patterns by Dewetting Aqueous Dispersions of Halloysite Nanotubes},
  journal = {The Journal of Physical Chemistry C},
  volume = {124},
  pages = {14},
  note = {doi:10.1021/acs.jpcc.0c01424.},
}

@article{Murray1926a,
  author = {Murray, C. D.},
  year = {1926},
  title = {The physiological principle of minimum work: I. The vascular system and the cost of blood volume},
  journal = {Proceedings of the National Academy of Sciences of the United States of America},
  volume = {12},
  pages = {207-214},
  note = {doi:10.1073/pnas.12.3.207.},
}

@article{Murray1926b,
  author = {Murray, C. D.},
  year = {1926},
  title = {The physiological principle of minimum work applied to the angle of branching of arteries},
  journal = {The Journal of general physiology},
  volume = {9},
  pages = {835-841},
  note = {doi:10.1085/jgp.9.6.835.},
}

@article{Strahler1957,
  author = {Strahler, A. N.},
  year = {1957},
  title = {Quantitative analysis of watershed geomorphology},
  journal = {Transactions American Geophysical Union},
  volume = {33},
  pages = {913-920},
  note = {doi:10.1029/TR038i006p00913.},
}

@book{weaire1999physics,
  title={The physics of foams},
  author={Weaire, Denis L and Hutzler, Stefan},
  year={1999},
  publisher={Oxford University Press}
}

@book{israelachvili2011intermolecular,
  title={Intermolecular and surface forces},
  author={Israelachvili, Jacob N},
  year={2011},
  publisher={Academic press}
}

@article{Zamir1982,
  author = {Zamir, M. and Medeiros, J. A.},
  year = {1982},
  title = {Arterial branching in man and monkey},
  journal = {J Gen Physiol},
  volume = {79},
  number = {3},
  month = {March},
  pages = {353-60},
  note = {doi:10.1085/jgp.79.3.353. PMID: 7077288. PMCID: PMC2215753},
}

@article{LaBarbera1990,
  title = {Principles of Design of Fluid Transport Systems in Zoology},
  volume = {249},
  ISSN = {1095-9203},
  url = {http://dx.doi.org/10.1126/science.2396104},
  DOI = {10.1126/science.2396104},
  number = {4972},
  journal = {Science},
  publisher = {American Association for the Advancement of Science (AAAS)},
  author = {LaBarbera,  Michael},
  year = {1990},
  month = aug,
  pages = {992–1000}
}

@inbook{Kizilova2015,
  title = {Biomimetic Composites Reinforced by Branched Nanofibers},
  ISBN = {9783319185439},
  ISSN = {1867-4941},
  url = {http://dx.doi.org/10.1007/978-3-319-18543-9_2},
  DOI = {10.1007/978-3-319-18543-9_2},
  booktitle = {Nanoplasmonics,  Nano-Optics,  Nanocomposites,  and Surface Studies},
  publisher = {Springer International Publishing},
  author = {Kizilova,  N.},
  year = {2015},
  pages = {7–23}
}

@article{Nunez-Iglesias_etal_2018,
    title = {A new {Python} library to analyse skeleton images confirms malaria parasite remodelling of the red blood cell membrane skeleton},
    volume = {6},
    issn = {2167-8359},
    url = {https://peerj.com/articles/4312},
    doi = {10.7717/peerj.4312},
    language = {en},
    urldate = {2026-07-29},
    journal = {PeerJ},
    publisher = {PeerJ Inc.},
    author = {Nunez-Iglesias, Juan and Blanch, Adam J. and Looker, Oliver and Dixon, Matthew W. and Tilley, Leann},
    month = feb,
    year = {2018},
    pages = {e4312},
}

@article{Arganda-Carreras_etal_2010,
    title = {{3D} reconstruction of histological sections: {Application} to mammary gland tissue},
    volume = {73},
    copyright = {Copyright © 2010 Wiley-Liss, Inc.},
    issn = {1097-0029},
    shorttitle = {{3D} reconstruction of histological sections},
    url = {https://onlinelibrary.wiley.com/doi/abs/10.1002/jemt.20829},
    doi = {10.1002/jemt.20829},
    language = {en},
    number = {11},
    urldate = {2026-08-05},
    journal = {Microscopy Research and Technique},
    author = {Arganda-Carreras, Ignacio and Fernández-González, Rodrigo and Muñoz-Barrutia, Arrate and Ortiz-De-Solorzano, Carlos},
    year = {2010},
    pages = {1019--1029},
}

@article{Hennig_etal_2026a,
    title = {Arc-{Length} {Characterization} of {Finite}, {Radial} {Growth} {Patterns}},
    issn = {0743-7463},
    url = {https://doi.org/10.1021/acs.langmuir.6c02052},
    doi = {10.1021/acs.langmuir.6c02052},
    urldate = {2026-08-05},
    journal = {Langmuir},
    volume = {42},
    number = {42},
    pages = {31193--31203},
    author = {Hennig, Andreas Andersen and Beechey-Newman, Ilaria and Kizilova, Natalya and Eiser, Erika},
    month = aug,
    year = {2026},
}

@article{Witten_Sander_1981,
    title = {Diffusion-{Limited} {Aggregation}, a {Kinetic} {Critical} {Phenomenon}},
    volume = {47},
    url = {https://link.aps.org/doi/10.1103/PhysRevLett.47.1400},
    doi = {10.1103/PhysRevLett.47.1400},
    number = {19},
    urldate = {2026-03-05},
    journal = {Physical Review Letters},
    publisher = {American Physical Society},
    author = {Witten, T. A. and Sander, L. M.},
    month = nov,
    year = {1981},
    pages = {1400--1403},
}

@book{Brochard-Wyart_etal_2019,
    title = {Essentials of {Soft} {Matter} {Science}},
    language = {en},
    author = {Brochard-Wyart, Françoise and Nassoy, Pierre and Puech, Pierre-Henri},
    year = {2019},
    publisher = {CRC Press},
    address = {Boca Raton},
}

@article{Devauchelle_etal_2012,
    title = {Ramification of stream networks},
    volume = {109},
    url = {https://www.pnas.org/doi/full/10.1073/pnas.1215218109},
    doi = {10.1073/pnas.1215218109},
    number = {51},
    urldate = {2026-08-19},
    journal = {Proceedings of the National Academy of Sciences},
    publisher = {Proceedings of the National Academy of Sciences},
    author = {Devauchelle, Olivier and Petroff, Alexander P. and Seybold, Hansjörg F. and Rothman, Daniel H.},
    month = dec,
    year = {2012},
    pages = {20832--20836},
}

@misc{hennig_2026_18954724,
  author = {Hennig, Andreas and Beechey-Newman, Ilaria and Kizilova, Natalya and Eiser, Erika},
  title = {Arc-length characterization of finite, radial growth patterns},
  month = mar,
  year = {2026},
  publisher = {Zenodo},
  doi = {10.5281/zenodo.18954724},
  url = {https://doi.org/10.5281/zenodo.18954724},
}

@article{Tsurusawa_Tanaka_2023,
    title = {Hierarchical amorphous ordering in colloidal gelation},
    volume = {19},
    issn = {1745-2473, 1745-2481},
    url = {https://www.nature.com/articles/s41567-023-02063-x},
    doi = {10.1038/s41567-023-02063-x},
    language = {en},
    number = {8},
    urldate = {2026-08-28},
    journal = {Nature Physics},
    author = {Tsurusawa, Hideyo and Tanaka, Hajime},
    month = aug,
    year = {2023},
    pages = {1171--1177},
}

@article{Lu_etal_2017,
    title = {Universal nanodroplet branches from confining the {Ouzo} effect},
    volume = {114},
    issn = {0027-8424, 1091-6490},
    url = {https://pnas.org/doi/full/10.1073/pnas.1704727114},
    doi = {10.1073/pnas.1704727114},
    language = {en},
    number = {39},
    urldate = {2026-08-30},
    journal = {Proceedings of the National Academy of Sciences},
    author = {Lu, Ziyang and Schaarsberg, Martin H. Klein and Zhu, Xiaojue and Yeo, Leslie Y. and Lohse, Detlef and Zhang, Xuehua},
    month = sep,
    year = {2017},
    pages = {10332--10337},
}
\bibliographystyle{rsc} 


\end{document}